\documentclass[aps,pre,twocolumn,superscriptaddress]{revtex4}
\usepackage{graphicx,color,appendix,amssymb,amsmath,hyperref}
\usepackage[british]{babel}

\begin{document}

\title{Ageing underdamped scaled Brownian motion: ensemble and time averaged particle displacements, non-ergodicity, and the failure of the overdamping approximation}

\author{Hadiseh Safdari}

\affiliation{Institute for Physics \& Astronomy, University of Potsdam, 14476 Potsdam-Golm, Germany}
\affiliation{Department of Physics, Shahid Beheshti University, 19839 Tehran, Iran}

\author{Andrey G. Cherstvy}
\affiliation{Institute for Physics \& Astronomy, University of Potsdam, 14476 Potsdam-Golm, Germany}

\author{Aleksei V. Chechkin}
\affiliation{Institute for Physics \& Astronomy, University of Potsdam, 14476 Potsdam-Golm, Germany}
\affiliation{Institute for Theoretical Physics, Kharkov Institute of Physics and Technology, 61108 Kharkov, Ukraine}
\affiliation{Department of Physics \& Astronomy, University of Padova, "Galileo Galilei" - DFA, 35131 Padova, Italy}

\author{Anna Bodrova}
\affiliation{Institute of Physics, Humboldt University Berlin, 12489 Berlin, Germany}
\affiliation{Faculty of Physics, M. V. Lomonosov Moscow State University, 119991 Moscow, Russia}

\author{Ralf Metzler}
\email{rmetzler@uni-potsdam.de}
\affiliation{Institute for Physics \& Astronomy, University of Potsdam, 14476 Potsdam-Golm, Germany}

\date{\today}

\begin{abstract}

We investigate both analytically and by computer simulations the ensemble averaged, time averaged, non-ergodic, and ageing properties of massive particles diffusing in a medium with a time dependent diffusivity. We call this stochastic diffusion process the (ageing) underdamped scaled Brownian motion (UDSBM). We demonstrate how the mean squared displacement (MSD) and the time averaged MSD of UDSBM are affected by the inertial term in the Langevin equation, both at short, intermediate, and even long diffusion times. In particular, we quantify the ballistic regime for the MSD and the time averaged MSD as well as the spread of individual time averaged MSD trajectories. One of the main effects we observe is that---both for the MSD and the time averaged MSD---for superdiffusive UDSBM the ballistic regime is much shorter than for ordinary Brownian motion. In contrast, for subdiffusive UDSBM the ballistic region extends to much longer diffusion times. Therefore, particular care needs to be taken when the overdamped limit indeed provides a correct description, even in the long time limit. We also analyze to what extent ergodicity in the Boltzmann-Khinchin sense in this non-stationary system is broken, both for subdiffusive and superdiffusive UDSBM. Finally, the limiting case of ultraslow UDSBM is considered, with a mixed logarithmic and power law dependence of the ensemble and time averaged MSDs of the particles. In the limit of strong ageing, remarkably, the ordinary UDSBM and the ultraslow UDSBM behave similarly in the short time ballistic limit. The approaches developed here open new ways for considering other stochastic processes under physically important conditions when a finite particle mass and ageing in the system cannot be neglected. 

\end{abstract}

\maketitle


\section{Introduction}

Anomalous diffusion processes feature a nonlinear growth of the ensemble  averaged mean squared displacement (MSD) of particles with time \cite{kehr87, bouc90,metz00, havl02,metz11,metz12, soko12, saxt12, fran13, bray13,metz14, soko15,wero15}, namely \begin{equation}\left\langle x^2(t) \right\rangle = \int_{-\infty}^{\infty}x^2 P(x,t) dx \sim  2 K_\alpha t^{\alpha}. \label{eq-ad} \end{equation} Here $P(x,t)$ is the probability density function (PDF) to find the tracer particle at time $t$ at position $x$ and $K_\alpha$ is the generalized diffusion coefficient with physical dimensions $ [K_\alpha]$= cm$^2$sec$^{-\alpha}$. The anomalous---and in general time local---scaling exponent $\alpha(t)$ distinguishes the regimes of subdiffusive ($0<\alpha<1$), normal ($\alpha=1$), and superdiffusive ($\alpha>1$) particle motions. The ballistic regime corresponds to $\alpha=2$. Hyperballistic MSD growth with $\alpha>2$ can occur, for instance, for particle diffusion in turbulent flows \cite{rich26, batc52} or for non-equilibrium initial conditions \cite{Hanggi}. 

Anomalous particle kinetics was detected in numerous physical and biophysical systems. From the perspective of crowded \cite{crow, greb16}  biological cells, the list of examples includes protein diffusion in living cells \cite{odde04, frad05, gold06,weis09}, motion of chromosomal loci \cite{gari09, webe10, wero12, huet15} and polymeric molecules \cite{he08}, diffusion of virus particles \cite{seis01}, motion of lipid and insulin granules inside cells \cite{odde11,tabe13}, diffusion of membrane lipids \cite{schw99a,kusu05, petr11a, petr11b, knel11, weis11, fran10, jeon12,sans13} and membrane-crowding proteins \cite{jeon13, vatt15, yama15, metz16}, dynamics of ion channels \cite{krap11,krap15a, krap15b,lape15} in biomembranes, diffusion of small molecules near cell membranes \cite{yama14, netz13, grub14}, active transport in cells \cite{sup1,sup2,sup3}, and, finally, the motion on the level of entire microorganisms \cite{Alves}.

In contrast to the universal Gaussian normal diffusion, anomalous diffusion processes are non-universal. There exists a variety of theoretical models sharing the same form (\ref{eq-ad}) of the MSD \cite{metz14}, including continuous time random walks describing diffusion with a divergent waiting time scale \cite{mont65, soko09, bark13} and trapping models in random energy landscapes \cite{traps}. In addition, models for particle motion in heterogeneous environments \cite{chec05, fuli13, cher13,cher14b, cher14c,cher15b, kaza16} and stochastic processes with distributed or time varying diffusion coefficient were considered \cite{slat14,lape14, cher16r}. Exponentially fast \cite{blac73, cher16, lube07} and logarithmically slow \cite{sina82, bray13, bodr15gg, bodr15, gode14, lomh13} anomalous diffusion processes are also worth mentioning here. Moreover, fractional Brownian motion and fractional Langevin equation motion with a power-law correlated noise \cite{bark09,gode16, grig16} can describe the dynamics of particles in viscoelastic media such as the cell cytoplasm.  Also, correlated continuous time random walks should be mentioned here \cite{metz10, chec09, magd12}. The adequate description of  some systems required the coupling of more than one anomalous diffusion mechanism \cite{krap11, tabe13, odde11}. 

Here we consider the remaining popular anomalous diffusion model, scaled Brownian motion (SBM) with the time dependent diffusion coefficient of the power law form \cite{muni02, jeon14, soko14,jeon15, safd15, bodr16}, \begin{equation} D(t) = \alpha K_\alpha t^{\alpha-1}. \label{eq-diff-sbm} \end{equation} SBM is a Gaussian and inherently non-stationary process. The power-law dependence (\ref{eq-diff-sbm}) of the particle diffusivity was widely used to describe i.a. subdiffusion in cellular fluids  \cite{kalla}, water diffusion in cells \cite{Latour}, and it naturally arises for the self-diffusion in granular gases \cite{Brilliantov, bril96, bodr16}. For more examples the reader is refereed to our recent study \cite{bodr16}. The exponent of $D(t)$ is $0< \alpha <1$ for subdiffusion,  $1<\alpha$ for superdiffusion, and $\alpha=0$ denotes ultraslow SBM diffusion (considered in Ref. \cite{bodr15}). 

Despite a great interest in the SBM process, the standard approaches usually deal with  \textit{massless} particles, the overdamped limit of the Langevin equation \cite{metz14}. The regime of underdamped motion---when the inertial term is non-negligible \cite{uhle30}---is typically less studied for anomalous diffusion processes. As exceptions we mention the fractional Langevin and fractional Klein-Kramers equations studied in Refs. \cite{klaf00, bark00}. In this case---under the conditions of weak coupling of particles to the thermal bath---the  ballistic diffusion is known to govern the short time dynamics \cite{Pavliotis, Hallerberg, Antczak}. Recently, the first results for the underdamped SBM (UDSBM) process were obtained by the authors in Ref. \cite{bodr16}. It was found \cite{bodr16} that for $\alpha>1$ the overdamped regime is reached rather soon, while for small positive $\alpha$ values an intermediate regime for the particle dynamics emerges and influences the particle dynamics, both for the MSD and the time averaged MSD. Finally, for ultraslow SBM at $\alpha=0$ the overdamped limit is not reached at all. Thus, a finite particle mass affects the dynamics at all time scales \cite{bodr16} and the description in terms of the conventional overdamped limit fails \cite{bodr15}. 

The current study clarifies which properties of UDSBM  \cite{bodr16} should be modified in the presence of ageing. The latter means that one starts recording the particle position after some ageing time $t_a$. In particular, for out-of-equilibrium processes such as SBM one expects severe effects of ageing onto the particle dynamics \cite{Henkel, Struik}. Therefore, the time interval $t_a$ impacts the statistical properties   \cite{metz14, Donth, henning}. Effects of ageing are observed, for instance, in glassy systems \cite{bonn05, ageing3,ageing4, ageing5,ageing6}, homogeneously cooled granular fluids \cite{Brey}, for diffusion in plasma cell membranes \cite{ageing0}, protein dynamics \cite{ageing1, ageing2}, in polymeric semiconductors \cite{neher}, as well as for blinking statistics of quantum dots \cite{ageing7, ageing8}.

We here generalize the stochastic SBM process to the underdamped and ageing situation. In Section \ref{sec-model} we introduce the observables and describe the routine for computer simulations. In Section \ref{sec-res} we present the main findings for the MSD, the time averaged MSD, and the ergodicity breaking parameter of ageing UDSBM. The cases of subdiffusion and superdiffusion are considered separately. We compare the results of analytical calculations and extensive computer simulations in different ageing regimes. In Section \ref{sec-uudsbm} the MSD and the time averaged MSD for the spacial case of ultraslow UDSBM are considered. In Section \ref{sec-disc} we discuss some applications of our results and conclude.

\section{Observables and Simulations Model}
\label{sec-model}

In addition to the standard characteristic of particle spreading given by the ensemble averaged MSD \cite{kehr87}, we are interested hereafter also in the the time averaged MSD. The latter is defined from a single particle trajectory $x(t)$ as \cite{metz14} \begin{equation} \label{eq-tamsd}\overline {\delta^2(\Delta)}= \frac{1}{T-\Delta}\int \limits_{0}^{T-\Delta} \left[x(t+ \Delta)-x(t)\right]^2 dt.\end{equation} The extension to higher dimensions is straightforward. Here, the lag time $\Delta$ is the width of the sliding window and $T$ is the total trajectory length. Expression (\ref{eq-tamsd}) is the standard measure to quantify particle displacements in single particle tracking experiments, when few but long time series $x(t)$ are typically available \cite{aust06}. It is complementary to the ensemble averaged MSD, widely used in the theoretical analysis of stochastic processes: there the averaging is performed at each time $t$ over the ensemble of $N$ given trajectories of the same length. When the measurement starts after time $t_a$ from the initiation of the process, the ageing time averaged MSD is  naturally defined as \cite{bark13} \begin{equation} \label{eq-tamsd-aged} \overline{\delta^2_a(\Delta)}= \frac{1}{T-\Delta} \int\limits_{t_a}^{T+t_a-\Delta} \left[x(t+\Delta)-x(t)\right]^2 dt.\end{equation} The average over $N$ realizations of the diffusion process yields the \textit{mean} time averaged MSD, \begin{equation} \left\langle \overline{\delta^2(\Delta)}\right \rangle = \frac{1}{N}\sum_{i=1}^N \overline{ \delta^2_i(\Delta)}, \label{eq-mean-tamsd} \end{equation} and analogously for $\left\langle \overline{\delta^2_a (\Delta)}\right \rangle$. This trajectory based averaging gives rise to a smoother variation of the time averaged MSD with the lag time, as compared to individual realizations,  (\ref{eq-tamsd}). Note that for stochastic processes with a pronounced \textit{scatter }of individual time averaged MSD realizations---such as continuous time random walks and heterogeneous diffusion processes \cite{metz14,cher13}---the determination of the mean (\ref{eq-mean-tamsd}) requires a substantial averaging sample to be generated \cite{metz14}.

For ergodic diffusion processes in the Boltzmann-Khinchin sense the MSD (\ref{eq-ad}) and the time averaged MSD (\ref{eq-tamsd}) coincide in the limit  $\Delta/T \ll 1$ \cite{metz14}. A quantitative measure of the ergodic properties of a stochastic process \cite{weak,weak2,soko08} is the ergodicity breaking parameter, EB, defined via the fourth moment of the time averaged MSD \cite{ryto87, bark09}, \begin{equation}\text{EB}(\Delta)=
\frac{\left\langle\left(\overline{\delta^2(\Delta)}\right)^2 \right\rangle-
\left\langle\overline{\delta^2(\Delta)}\right\rangle^2}{\left\langle\overline{ \delta^2(\Delta)} \right\rangle^2}= \left\langle\xi^2(\Delta)\right\rangle-1. \label{eq-eb-general}\end{equation} Here, the ratio \begin{equation}\xi(\Delta)= \frac{ \overline{\delta^2(\Delta)}}{\left\langle\overline{\delta^2(\Delta)} \right\rangle} \end{equation} is a dimensionless parameter that quantifies the relative deviation \cite{akim15} of individual time averaged MSDs  about their mean. The characteristics employing the higher moments such as skewness and kurtosis can be implemented additionally to the EB parameter, to characterize finer details of the spread of time averaged MSD trajectories \cite{gode16}.

For SBM considered herein, we numerically solve the stochastic Langevin equation for \textit{massive} particles \cite{bodr16}
 \begin{equation} \frac{d^2x(t)}{dt^2}+\gamma(t)\frac{dx(t)}{dt}=\sqrt{2D(t)} \gamma(t)\eta(t), \label{eq-under} \end{equation} driven by the Gaussian noise $\eta(t)$ with zero mean $\left\langle \eta(t) \right \rangle=0$ and unit variance $\left\langle \eta(t)\eta(t') \right \rangle=\delta(t-t').$
The friction coefficient is a time dependent function, $\gamma(t)=\tau_{v}^{-1}(t)$, where \begin{equation} \tau_{v}^{-1}(t)={\gamma_0} \sqrt{\frac{\mathcal{T}(t)} {\mathcal{T}(0)}} \end{equation} contains the time dependent temperature $\mathcal{T}(t)$. For instance, for force-free cooling granular gases this dependence is characterized by the law  \cite{bodr16} \begin{equation} \mathcal{T}(t)= \frac{\mathcal{T}_{0}}{\left(1+t/\tau_0\right)^{2-2\alpha}}. \label{eq-temp-variations} \end{equation} For viscoelastic granular gases the exponent is $\alpha=1/6$, while for granular gases with a constant restitution coefficient $\alpha=0$, see Refs. \cite{Brilliantov,bodr15gg,bril96}. Correspondingly, the time dependent diffusion coefficient of the particles is \begin{equation} D(t)=\frac{D_0}{\left(1+t/ \tau_0\right)^{1-\alpha}}, \end{equation} where the initial values are $\mathcal{T}_{0} =\mathcal{T}(0)$ and $D_0 = D(0).$ Putting the Boltzmann constant hereafter to unity ($k_B=1$) we get the time-local fluctuation-dissipation relation \cite{bodr16} \begin{equation}D(t)= \frac{\mathcal{T}(t)} {\gamma(t)m}. \end{equation} Note that the characteristic scale of the temperature variation, $\tau_0$, is much longer than the typical relaxation time in the system,  \begin{equation} \tau_0\gamma_0\gg1. \label{eq-tau0-gamma0-big} \end{equation} This condition assures the applicability of the initial Langevin equation (\ref{eq-under}). 

The second order Langevin equation (\ref{eq-under}) is equivalent to two differential equations of the first order for the increments of the particle position $x(t)$ and velocity $v(t)$ \cite{stoch1, stoch2}, namely (assuming unit particle mass $m=1$ from hereon) \begin{align} d v(t) =\sqrt{2 D(t)}\gamma(t) \eta(t) \sqrt{dt}-\gamma(t) v(t) dt, \label{sde-set-1} \end{align} \begin{align} dx(t) =v(t) dt. \label{sde-set-2} \end{align} We discretize this system of equations in $T/\delta t$ steps and use the unit time step in our simulations ($\delta t=1$). Hence, on time step $t_{n+1}$ the following discrete scheme is solved \begin{align}  \nonumber v(t_{n+1}) = & v(t_{n})+\sqrt{2 D(t_n)}\gamma(t_n) \eta(t_n) \sqrt{t_{n+1}-t_n} \\ & -\gamma(t_n) v(t_n) (t_{n+1}-t_n), \end{align}   \begin{eqnarray}  x(t_{n+1}) =x(t_{n})+v(t_n)(t_{n+1}-t_n). \label{sde-set-dis} \end{eqnarray}

\section{Main results: ageing UDSBM}
\label{sec-res}

In this section we present our results for the ensemble and time averaged MSDs of UDSBM. We also quantify the amplitude scatter of individual time averaged MSD trajectories of this process. We first present the analytical results for the UDSBM process and then compare them with computer simulations. 

\subsection{MSD}

To obtain the ensemble averaged MSD, we start with the velocity-velocity correlation function, that can be directly obtained via integration of the Langevin equation (\ref{eq-under}): assuming without loss of generality that $t_2>t_1$ and applying the same approximations as described in Ref. \cite{bodr16} to evaluate the integrals, we find \begin{align} \nonumber \left\langle v(t_1)v(t_2) \right\rangle \approx &  D_0 \gamma_0 \left(1+\frac{t_1} {\tau_0} \right)^{2\alpha-2}  \\ \times & \exp \left\{\frac{\tau_0 \gamma_0}{\alpha}\left[\left(1+\frac{t_1} {\tau_0}\right)^{\alpha}- \left(1+ \frac{t_2}{\tau_0}\right)^{\alpha} \right]\right\}. \label{v_corr} \end{align} The correlation function (\ref{v_corr}) is obtained under the condition  The ensemble averaged MSD of diffusing particles can be obtained via the integration of the correlation function (\ref{v_corr}),  \begin{align} \label{x_corr} &\left\langle x^2(t) \right\rangle = 2\int_{0}^{t}dt_1 \int_{t_1}^{t} dt_2\left\langle v(t_1)v(t_2) \right \rangle. \end{align} The reader is referred to our recent study \cite{bodr16} for details on the derivation of $\left< v(t_1)v(t_2) \right>$ and the MSD for the non-ageing UDSBM process. In short, at $t_a=0$ one gets \begin{align} \label{MSD_nonaged} \nonumber \left\langle x^2(t) \right\rangle \approx & 2D_0\left\{ \frac{\tau_0}{\alpha} \left(\left(1+\frac{t}{\tau_0} \right)^{\alpha}-1\right)  \right. \\ &+\left. \gamma_0^{-1} \left(\exp\left(-\frac{\tau_0 \gamma_0} {\alpha}\left[\left(1+ \frac{t}{\tau_0} \right)^{\alpha}-1\right] \right)-1 \right) \right\}\end{align}

In the more general situation when the recording of particle position starts after the ageing time $t_a,$ the MSD of the ageing UDSBM process is described by \begin{align} \label{eq-general-msd}\left\langle x_a^2(t) \right\rangle =2\int_{t_a}^{t_a+t}dt_1
\int_{t_1}^{t_a+t}dt_2\left\langle v(t_1)v(t_2) \right \rangle. \end{align} The integration over $t_2$ can be performed to yield \begin{align} \label{eq-general-msd-via-gamma} \nonumber \left\langle x_a^2(t) \right\rangle = & \\ \nonumber &\frac{2 D_0 \tau_0 \gamma_0} {\alpha}\left(\frac{\alpha}{\tau_0 \gamma_0} \right)^{\frac{1}{\alpha}}
\int_{t_a}^{t_a+t}dt_1 \left(1+\frac{t_1}{\tau_0} \right)^{2\alpha-2}  \\
& \nonumber\times\exp\left(\frac{\tau_0 \gamma_0}{\alpha}\left[\left(1+\frac{t_1}{\tau_0} \right)^{\alpha}\right]\right) \\ \nonumber &
\times \left\{ \Gamma\left(\frac{1}{\alpha},\frac{\tau_0 \gamma_0}{\alpha} \left[1+\frac{t_1}{\tau_0}\right]^{\alpha}\right) \right. \\  &\left.-
\Gamma\left(\frac{1}{\alpha},\frac{\tau_0 \gamma_0 }{\alpha}\left[1+\frac{t+t_a} {\tau_0}\right]^{\alpha}\right)\right\}, \end{align} where $\Gamma(n,x)$ is the generalized Gamma function \cite{abra72}. Since throughout this paper we limit ourselves to the regime (\ref{eq-tau0-gamma0-big}) and the values of the scaling exponents $\alpha$ are not very large ($|\alpha|\lesssim2-5$) for most realistic applications, the expansion of the Gamma functions for large arguments can be performed. The same level of approximations was implemented when obtaining the velocity-velocity correlation function in Eq.  (\ref{v_corr}). After expansion (up to the appropriate order) this procedure yields the following result for the MSD, \begin{align} \nonumber \left\langle x_a^2(t) 
\right\rangle \approx & \frac{2 D_0 \tau_0}{\alpha}
\left[\left(1+\frac{t+t_a}{\tau_0}\right)^{\alpha}- \left(1+\frac{t_a}{\tau_0} \right)^{\alpha} \right] \\ \nonumber & + \frac{2 D_0}{\gamma_0} \left[\exp \left(-\frac{\tau_0 \gamma_0} {\alpha}\left[\left(1+\frac{t_a+t}{\tau_0}\right)^{\alpha} \right.\right.\right. \\ & \left.\left.\left.-\left(1+\frac{t_a}{\tau_0}\right)^{\alpha}  \right]\right)-1 \right].\label{eq-msd-general-after-G-expansion} \end{align}  

\subsubsection{Limiting cases}

\begin{figure} 
\includegraphics[width=7cm]{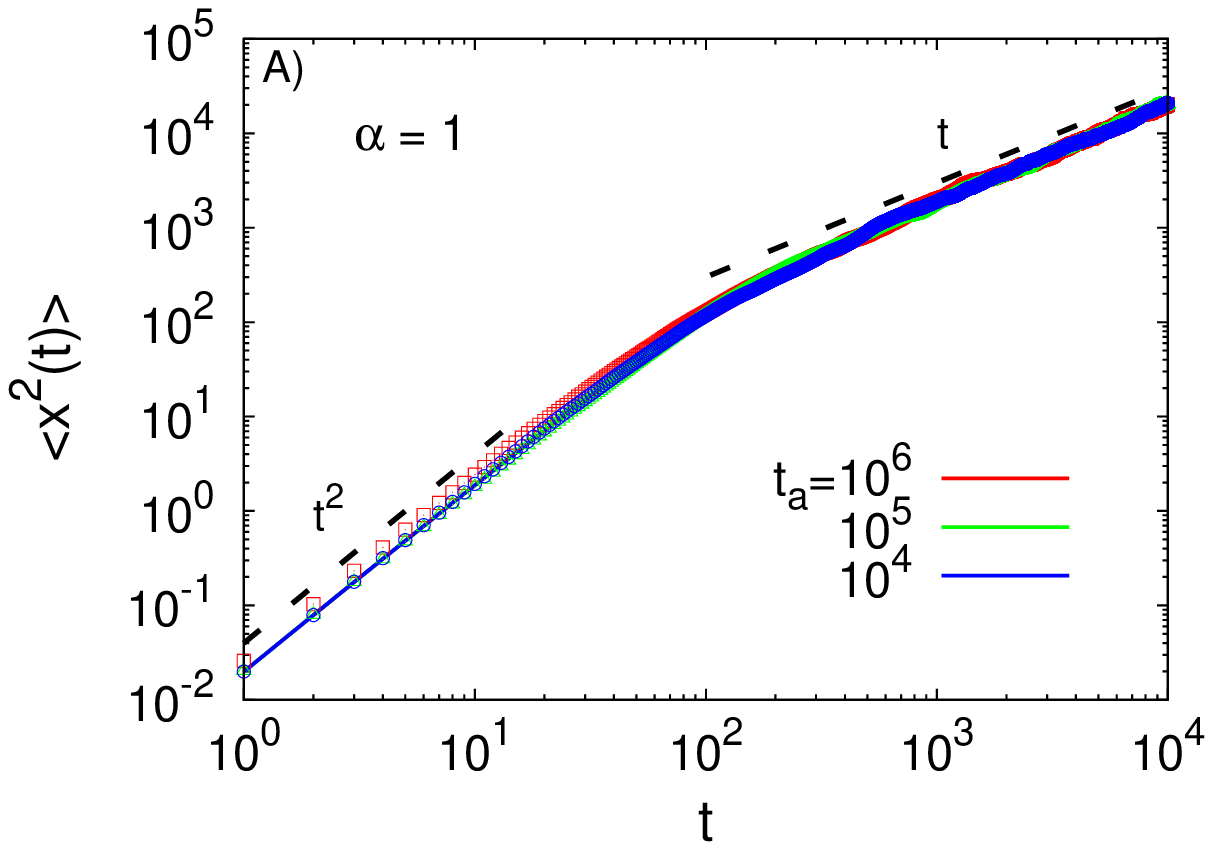} 
\includegraphics[width=7cm]{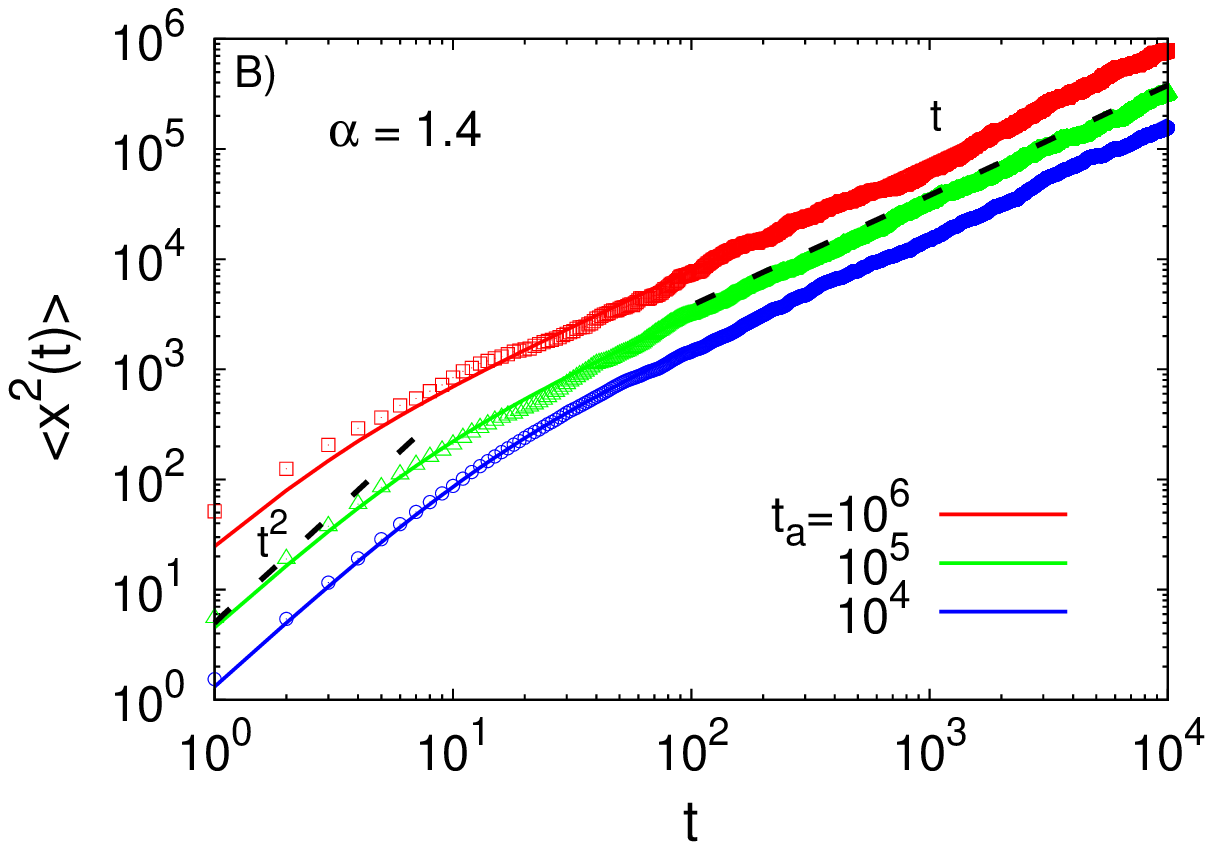}
\includegraphics[width=7cm]{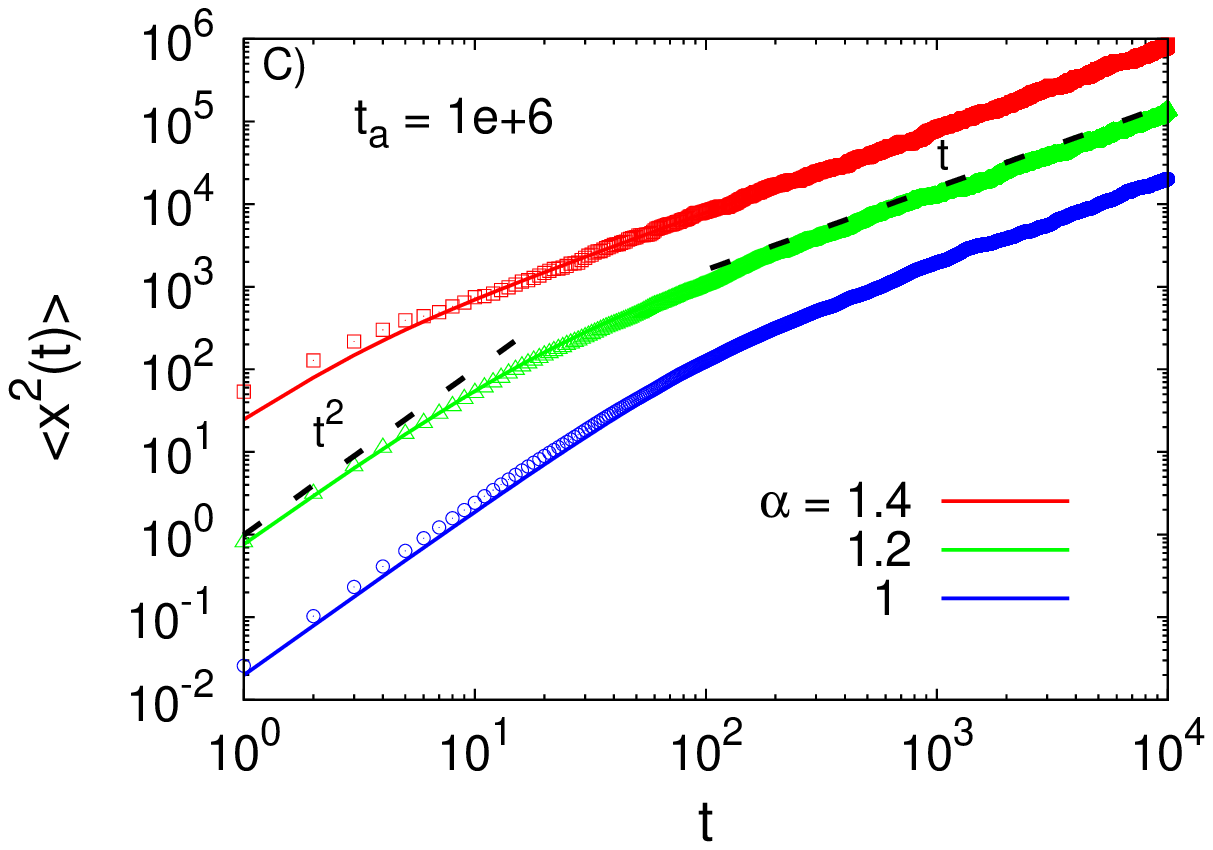}  
\caption{MSD of the ageing UDSBM process: analytical results (solid lines,
with the full expression of Eq. (\ref{eq-msd-general-after-G-expansion})
and with the asymptotes of Eqs. (\ref{ballisticMSD}) and (\ref{normal-MSD})
shown) and the results of our computer simulations (data points), plotted for
$\alpha=1$  (panel A) and $\alpha= 1.4$ (panel B) and different ageing times
$t_a$. Parameters: $T=10^{4}$, $\tau_0=100$, $\gamma_0=0.02$. Panel C shows
the MSD for the case of strong ageing  $t_a=10^6\gg T$ and different diffusion
exponents. For superdiffusive UDSBM, the ballistic MSD regime shrinks to
shorter times $t\ll1/\gamma_0$ for larger $\alpha$ values, in accord with the
analytical prediction (\ref{tmin}).} \label{fig-msd-supperdiff}
\end{figure}

Let us mention here some special cases. If we put $t_a=0$ in Eq. (\ref{eq-msd-general-after-G-expansion}), the MSD of the non-aged UDSBM process (\ref{MSD_nonaged}) is recovered. For  $\alpha=1$ the MSD behaves as that for standard Brownian motion \cite{uhle30}, namely \begin{align} \left\langle x_a^2(t) \right\rangle=\left<x^2( t)\right> = 2D_0\left[ t-\gamma_0^{-1}\left(1-\exp(-\gamma_0 t)\right)\right].\end{align} This expression and its scaling behaviors are shown in Fig. \ref{fig-msd-supperdiff}A. 

In the limit of very long observation and long ageing times, when $\tau_0 \ll t_a \ll t$, we can neglect the second square bracket term in Eq. (\ref{eq-msd-general-after-G-expansion}). The final MSD then coincides with the MSD of non-ageing UDSBM at long observation times  \cite{bodr16}, namely \begin{align} \label{eq-weak-ageing} \left\langle x_a^2(t) \right\rangle \sim  \frac{2 D_0 \tau_0 }{\alpha} \left(\frac{t}{\tau_0} \right)^{\alpha} \simeq t^\alpha.\end{align}

The most interesting situation emerges when the ageing time is the longest time scale in the problem, $\gamma_0^{-1} \ll \tau_0\ll t_a$ and $t\ll t_a$. For very small values of the argument (arg) of the exponential function in Eq. (\ref{eq-msd-general-after-G-expansion}), \begin{align} \text{arg}= \frac{\tau_0 \gamma_0}{\alpha}\left[\left(1+\frac{t_a+t} {\tau_0}\right)^{\alpha}- \left(1+\frac{t_a} {\tau_0}\right)^{\alpha}  \right] \ll1, \label{domain_S}\end{align} after expanding Eq. (\ref{eq-msd-general-after-G-expansion}) the ageing MSD shows the initial ballistic growth regime, \begin{align} \left\langle x_a^2(t) \right\rangle \sim D_0 \gamma_0\left(\frac{t_a}{\tau_0} \right)^{2 \alpha-2}t^2 \simeq t^2. \label{ballisticMSD} \end{align} Conversely, in the limit arg$\gg1$ expression (\ref{eq-msd-general-after-G-expansion}) yields the normal diffusion regime, \begin{align} \left\langle x_a^2(t) \right\rangle \sim 2D_0 t\left\{ \begin{aligned}
&\left(\frac{t_a}{\tau_0}\right)^{\alpha-1} ,~~  \alpha > 1\\
&\left(\frac{\tau_0}{t_a}\right)^{1-\alpha},~~  \alpha < 1
\end{aligned}.  \right.    \label{normal-MSD} \end{align}
This long ageing time MSD behavior is equivalent to that of ageing SBM considered in Ref. \cite{jeon15} and it features a linear dependence on the diffusion time $t$. Therefore, \textit{no anomalous diffusion regime at all} is observed for the UDSBM processes when the ageing time is the longest time scale in the problem, see Fig. \ref{fig-msd-supperdiff}. This is a quite remarkable effect of strong ageing. 

For comparison, note that for the non-ageing situation the MSD asymptotes for the initial, intermediate, and long time behaviors of the MSD of UDSBM are, respectively, \begin{equation}\left\langle x^2(t) \right\rangle \sim D_0 \gamma_0 t^2, \label{eq-d0-bm} \end{equation} \begin{equation}\left\langle x^2(t) \right\rangle \sim 2 D_0 t,\label{eq-linear-brownian-scaling} \end{equation} and $\left\langle x^2(t) \right\rangle \sim D_0\tau_0\alpha^{-1}(t/\tau_0)^{\alpha}$, as derived in Ref. \cite{bodr16}.

\begin{figure}
\includegraphics[width=7cm]{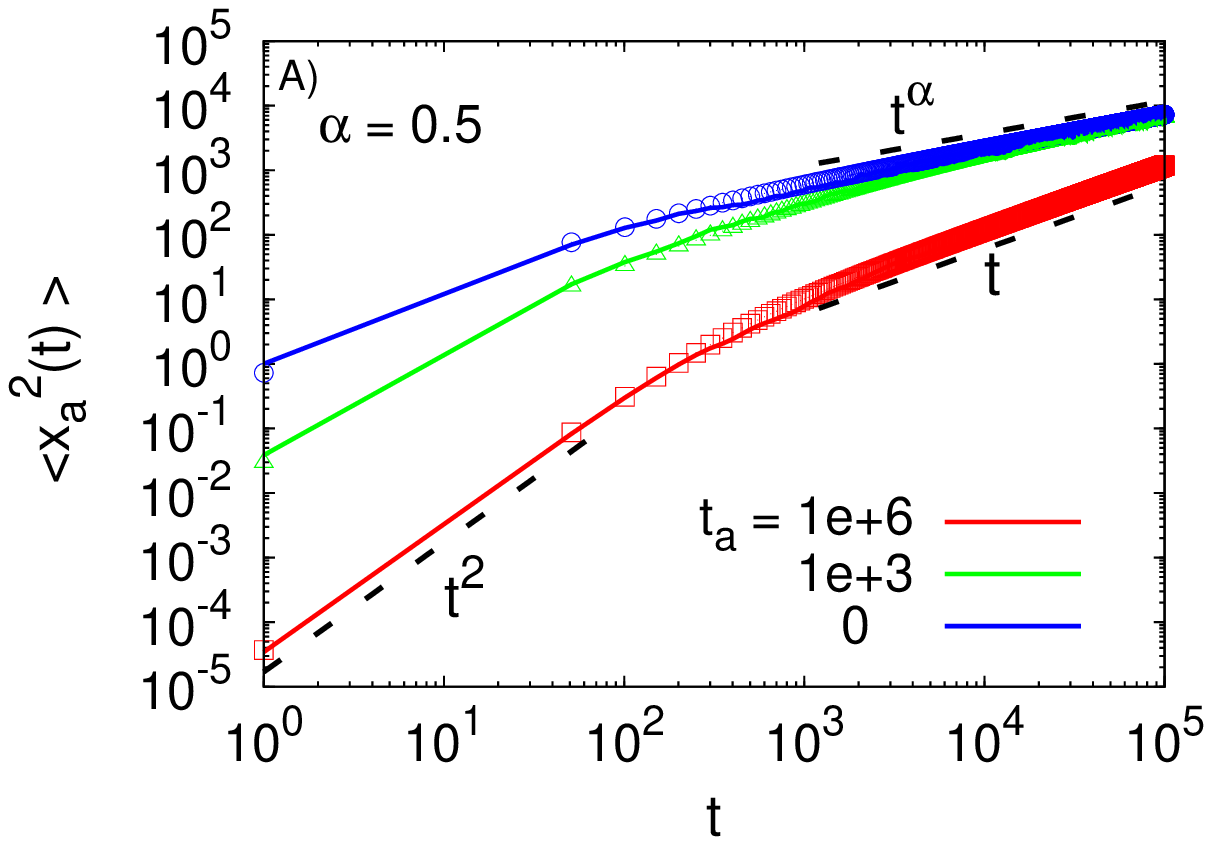}
\includegraphics[width=7cm]{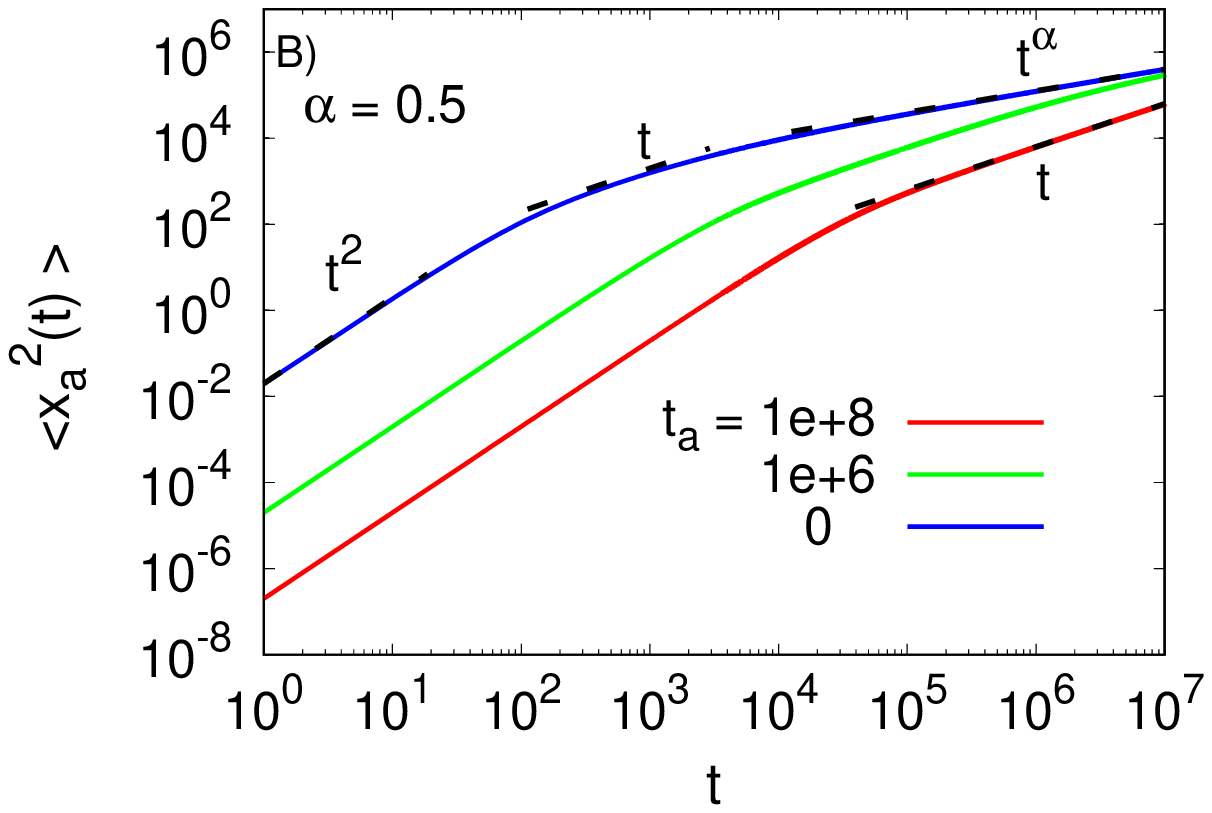}
\caption{Delay of the overdamping transition for ageing UDSBM. (A) Theoretical results (solid lines) and computer simulations (data points) results for the MSD of the ageing UDSBM process, obtained for $\alpha=0.5$. The initial ballistic regime is the dashed line given by Eq. (\ref{ballisticMSD}). The intermediate normal diffusion asymptote is the dashed line according to Eq. (\ref{normal-MSD}). The long time anomalous MSD asymptote follows Eq. (\ref{eq-msd-anomalous-long-time}). The trace length is $T=10^5$ and the ageing times $t_a$ are indicated in the plots. Parameters:  $D_0=1$, $\gamma_0=1$, $\tau_0=30$ and $m=1$. (B) Theoretical results for the MSD, plotted for different ageing times and for $\alpha=0.5,$ $D_0=1$, $\gamma_0=0.02$, $\tau_0=10^3$, $m=1,$ and  $T=10^7$, show three different regimes of the MSD scaling. Note that the values of $\tau_0$, $\gamma_0$, and $T$ in panel B differ from those in panel A. The dashed asymptotes for the ageing UDSBM are according to Eqs. (\ref{ballisticMSD}), (\ref{normal-MSD}), and (\ref{eq-msd-anomalous-long-time}).} \label{fig-msd} \end{figure}

\subsubsection{Superdiffusion versus subdiffusion}

Different effects in the particle dynamics take place for superdiffusive as compared to subdiffusive scaling exponents, as we show here. For $\alpha > 1$ and at $ \gamma_0^{-1} \ll \tau_0\ll t_a$ (strong ageing limit), the condition of arg$\gg1 $ is satisfied for all diffusion times, starting from very short times  \begin{equation}t_\text{min} \sim \gamma_0^{-1} \left(\frac {\tau_0}{t_a}\right)^{\alpha-1}\ll\gamma_0^{-1}\ll\tau_0 \ll t_a. \label{tmin} \end{equation} The ballistic regime (\ref{ballisticMSD}) is observed for  $t< t_\text{min}$. In other words, almost for the entire observation interval \textit{{normal }}diffusion is observed, yet the effective diffusion constant becomes much larger than in the non-ageing situation (\ref{eq-d0-bm}), that is  \begin{equation} D_\text{eff}(t_a) =D_0 \gamma_0 \left(\frac{t_a}{\tau_0} \right)^{\alpha-1} \gg D_0 \gamma_0,\end{equation} as follows from Eq. (\ref{normal-MSD}). The short initial region of ballistic diffusion and the long domain of normal diffusion  for the situation $\alpha>1$ are clearly visible in Fig. \ref{fig-msd-supperdiff}B,C. It is also seen that the region of normal diffusion extends towards shorter times with growing ageing times and thus the region of ballistic diffusion shrinks. This trend agrees with the estimate (\ref{tmin}) for $t_\text{min}$. This is another a priori surprising behavior of ageing UDSBM.

For subdiffusive exponents $0<\alpha< 1$ of the UDSBM process the condition  (\ref{domain_S}) is satisfied for much longer observation times, namely \begin{equation}t < t_\text{min}\sim \gamma_0^{-1}\left(\frac{t_a}{\tau_0} \right)^{1-\alpha} \gg\gamma_0^{-1} \label{tmin2} \end{equation} Hence, the ballistic diffusion regime  (\ref{ballisticMSD}) extends for times much longer than the relaxation time for normal diffusion, $\gamma_0^{-1}$. Thus, this ballistic regime for subdiffusive UDSBM persists even much longer than that for superdiffusive situation. However, the effective diffusion constant following from Eq. (\ref{ballisticMSD}), \begin{equation}D_\text{eff}(t_a) = D_0\gamma_0\left(\frac{ \tau_0}{t_a} \right)^{2-2\alpha},\end{equation} becomes much smaller than the basal value $D_0\gamma_0$ in Eq. (\ref{eq-d0-bm}). 

The MSD behavior for subdiffusive UDSBM is illustrated in Fig. \ref{fig-msd} for two sets of the model parameters. Since the time $t_\text{min}$ in Eq. (\ref{tmin2}) grows with the ageing time $t_a$, the region of ballistic diffusion becomes more extended, as clearly seen when comparing the curves for different $t_a$ values in Fig. \ref{fig-msd}. The MSD reveals a good agreement of theory and computer simulations, for all values of the model parameters examined. In Fig. \ref{fig-msd} we show the MSD for the trace lengths  $T=10^{5}$ and $T=10^{7}$ on panels A and B, respectively. We observe that for long $t_a$ the region of initial ballistic diffusion and the intermediate regime of normal diffusion shift towards longer times. For subdiffusive realizations, in the long time limit $t\gg t_a$ the anomalous behavior \begin{equation}\left\langle x_a^2(t) \right\rangle \sim \frac{2D_0\tau_0}{\alpha} \left(\frac{t}{\tau_0}\right) ^\alpha \simeq t^{\alpha} \label{eq-msd-anomalous-long-time} \end{equation} persists, as follows from Eq. (\ref{eq-msd-general-after-G-expansion}). In Fig. \ref{fig-msd}, however, this regime is realized for $t_a=0$ only because of a relatively short trajectory length $T$. For the case of superdiffusion in Fig. \ref{fig-msd-supperdiff}, this anomalous regime is not visible at all because the values of the ageing time used are large compared to the trace length $T$.   
 
\subsection{Time averaged MSD}

\subsubsection{General expressions}

The time averaged MSD (\ref{eq-tamsd-aged}) for the ageing UDSBM process is defined as \cite{metz14} \begin{align} \label{eq-general-tamasd} \nonumber \left\langle \overline{\delta^2_a(\Delta)} \right\rangle=&\frac{1}{T-\Delta} \int_{t_a}^{T+t_a-\Delta} dt  \\ & \times\left[\left\langle  x^2(t+\Delta)\right \rangle-\left\langle  x^2(t)\right\rangle -2A(t,\Delta)\right],\end{align} where the last term is computed from the non-ageing velocity-velocity correlation function (\ref{v_corr}) as \begin{equation} A(t,\Delta)=\int_0^{t}dt_1 \int_{t}^{t+ \Delta}dt_2 \left\langle v(t_1)v(t_2) \right\rangle.\end{equation} Following the strategy outlined in Ref. \cite{bodr16}, we divide the integral in Eq. (\ref{eq-general-tamasd}) {formally} into two parts, namely \begin{align} \left\langle\overline{\delta^2_a(\Delta)}\right\rangle = \left\langle\overline{ \delta_{0,a}^2(\Delta)}\right\rangle+ \left\langle \Xi_a(\Delta)\right\rangle. \label{eq-split}\end{align}  The first term here corresponds to the time averaged MSD of the ageing SBM process for massless particles \cite{safd15,jeon15}, \begin{align} \label{eq-tamsd-no-ageing-massive-delta0}
\nonumber \left\langle\overline{\delta_{0,a}^2(\Delta)}\right\rangle
=& \frac{2 D_0 \tau_0^{2}}{\alpha (1+\alpha)(T-\Delta)}  \\ \nonumber & \times \left[\left(1+\frac{T+t_a} {\tau_0}\right)^{\alpha+1}-
\left(1+\frac{t_a+\Delta}{\tau_0}\right)^{\alpha+1}\right. \nonumber \\ & -\left.\left(1+\frac{T+t_a-\Delta}{\tau_0}\right)^{\alpha+1}+ \left(1+\frac{t_a} {\tau_0}\right)^{\alpha+1}\right], \end{align} corresponding to the overdamped limit of the process \cite{safd15,jeon15}. The second term $\left\langle \Xi_a(\Delta)\right\rangle$ is due to the inertial term in the Langevin equation, which is absent in the standard SBM process. Under the condition (\ref{eq-tau0-gamma0-big}) we obtain the closed form solution
\begin{align}
\label{tamsd-SP1}
\left\langle \Xi_a(\Delta)\right\rangle \approx & \frac{2 D_0}{\gamma_0
(T-\Delta)} \int_{t_a}^{T+t_a-\Delta}dt\\
\nonumber & \times \left\{\exp\left(- \frac{\tau_0\gamma_0}{\alpha}\left[\left(
1+\frac{t+\Delta}{\tau_0}\right)^{\alpha} \right.\right.\right. \\
\nonumber & \left.\left.\left.-\left( 1+\frac{t}{\tau_0}\right)^{\alpha}\right]
\right)-1 \right\}.
\end{align}
The final integration cannot be performed for arbitrary values of $\alpha$. Below,
we consider the important limiting cases.

\subsubsection{Limiting cases}

At $t_a\to0$ we recover the time averaged MSD of the non-ageing UDSBM process,
see Eqs. (42), (43) in Ref. \cite{bodr16}.  For normal diffusion at $\alpha=1$
from Eqs. (\ref{eq-split}), (\ref{eq-tamsd-no-ageing-massive-delta0}),
and (\ref{tamsd-SP1}) we observe, as expected, no dependence
on the ageing time,
\begin{eqnarray}
\label{eq-general-tamasd-3}
\left\langle\overline{\delta^2_a(\Delta)}\right\rangle&=&
\left<{x^2_a(\Delta)}\right> = \left<{x^2(\Delta)}\right>\\
\nonumber &=& 2D_0\left[\Delta-\gamma_0^{-1}\left(1-\exp(-\gamma_0
\Delta)\right)\right].
\end{eqnarray}
In the most interesting limit of strong
ageing, when the condition
\begin{equation}
t_a \gg T \gg \{\Delta,\tau_0\}
\label{eq-strong-ageing-condition}
\end{equation}
is satisfied, the leading
order term in Eq. (\ref{eq-tamsd-no-ageing-massive-delta0}) grows linearly with
the lag time,
\begin{align}
\left\langle \overline{\delta_{0,a}^2 (\Delta)}
\right\rangle \sim 2 D_0 \Delta \left(\frac{t_a} {\tau_0}\right)^{\alpha-1}
\simeq \Delta.
\label{tamsd-asimp1}
\end{align}
In the same limit
Eq. (\ref{tamsd-SP1}) can be represented by
\begin{eqnarray}
\label{tamsd-SP4}
\nonumber
\left\langle\Xi_a(\Delta)\right\rangle\approx-\frac{2D_0}{\gamma_0}\left\{1-
\frac{1}{T-\Delta}\int_{t_a}^{T+t_a-\Delta}dt\right.\\
\times \left.   \exp\left[-\gamma_0 \Delta\left(\frac{t}
{\tau_0}\right)^{\alpha-1}\right]\right\}.
\end{eqnarray}
This expression will
be used below, for instance, to estimate the time intervals for the initial
ballistic behavior of the time averaged MSD of ageing UDSBM. We consider the
limit of strong ageing in Sec. \ref{sec-strong-ageing}, while the limit of
short ageing times is presented in Sec. \ref{sec-weak-ageing}, for the sake
of completeness. Note also that for non-ageing UDSBM in the intermediate lag
time regime  $\tau_0\ll\Delta\ll T$ the leading scaling for the time averaged
MSD is linear, similar to that of the overdamped SBM process  \cite{bodr16},
\begin{align}
\left\langle \overline{\delta_{0}^2(\Delta)} \right\rangle
\sim \frac{2 D_0 \Delta}{\alpha} \left(\frac{T} {\tau_0}\right)^{\alpha-1}
\simeq \Delta.
\label{eq-tamsd-anna-eq17}
\end{align}

\subsubsection{Superdiffusion versus subdiffusion: strong ageing}

\label{sec-strong-ageing}

\begin{figure} \includegraphics[width=7cm]{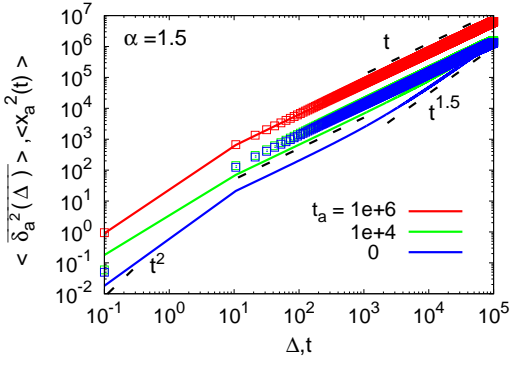}
\caption{Analytical results for the MSD (solid lines) and time averaged MSD (points) of ageing superdiffusive UDSBM at  $\alpha=1.5$. The asymptotes for the initial ballistic, intermediate linear, and long time anomalous behavior are according to Eqs. (\ref{ballisticMSD}), (\ref{normal-MSD}) and (\ref{eq-weak-ageing}), respectively. Note that long ageing times diminish and eventually remove the weak ergodicity breaking for ageing UDSBM. Parameters: $T=10^5$, $\tau_0=10^3$, and $\gamma_0=0.02$.} \label{fig-MSD_150} \end{figure}

\begin{figure} 
\includegraphics[width=7cm]{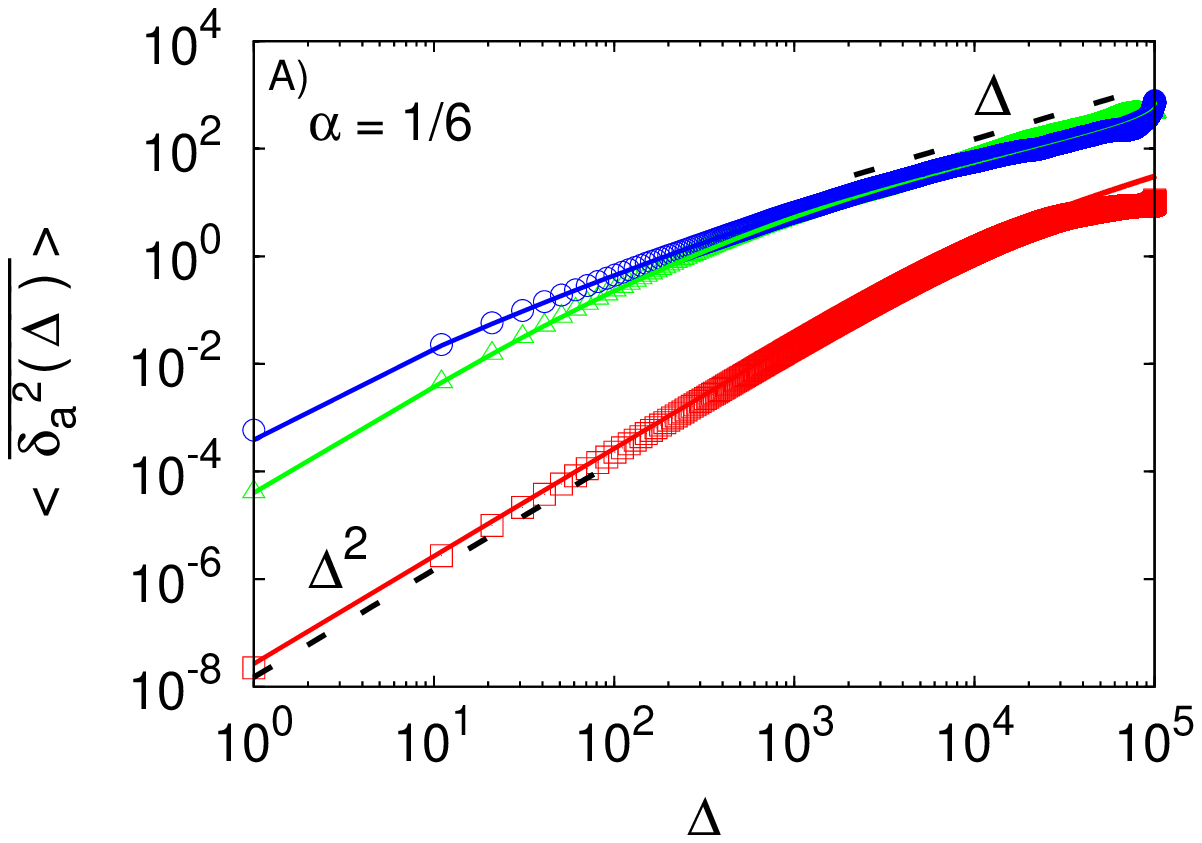} 
\includegraphics[width=7cm]{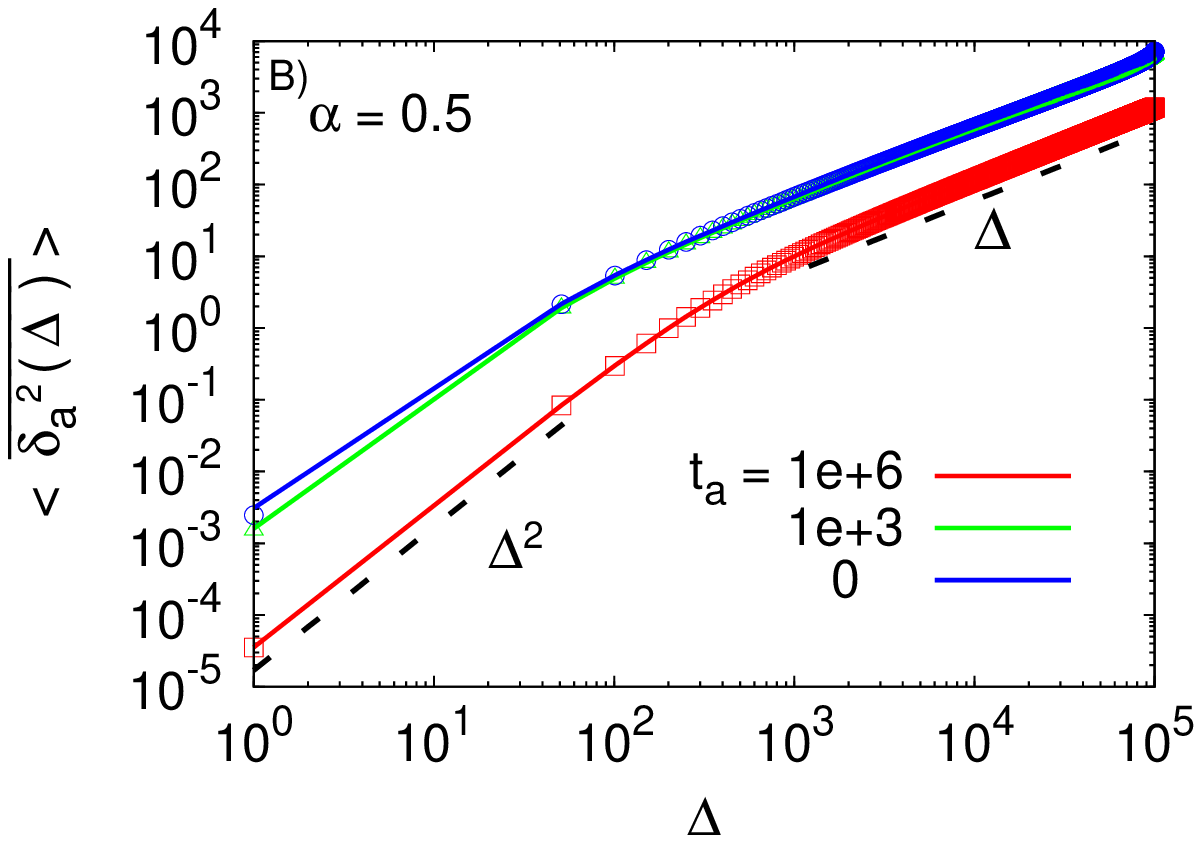} 
\caption{Theoretical results (solid curves, Eqs.  (\ref{eq-tamsd-no-ageing-massive-delta0}) and (\ref{tamsd-SP1})) and the results of computer simulations (data points) for the time averaged MSD of ageing UDSBM. The asymptotes shown as the dashed lines are according to  Eqs.  (\ref{eq-strong-ageing-limit-tamsd}) and Eqs. (\ref{tamsd-asimp1}), (\ref{eq-tamsd-anna-eq17}) for short and intermediate lag times, respectively. The findings are plotted for $\alpha=1/6$ (panel A) and $\alpha=$1/2 (panel B), with  $T=10^{5}$, $\tau_0=10^2$, and $\gamma_0=0.1$.}
\label{fig-tamsd-ballistic}\end{figure}

We start our analysis with the case of superdiffusion, presented in Fig. \ref{fig-MSD_150}. As one can see, the argument of the exponential function in Eq. (\ref{tamsd-SP4}) in the limit of strong ageing becomes very large already for lag times much shorter than the characteristic relaxation time, $1/\gamma_0$. The contribution of  the term $\left\langle \Xi_a(\Delta)\right\rangle$ to the time averaged MSD (\ref{eq-split}) can then be neglected, as compared to the leading Brownian term given by Eq. (\ref{tamsd-asimp1}). The initial ballistic regime in the time averaged MSD in this limit $t_a\gg T$ is then observed only for very short times $\Delta\ll 1/\gamma_0$, as indeed demonstrated in Fig. \ref{fig-MSD_150}. 

Here we observe an interesting effect, namely, with increasing lag times the MSD scaling exponent changes from the ballistic value of $\alpha=$2 to the normal diffusion value $\alpha=$1 and then back to a higher value of $\alpha=$1.5. The reader is also referred to Fig. 1 of Ref. \cite{bodr16} for the behavior of non-ageing UDSBM processes. Also, it is important to note that for long ageing times the ageing UDSBM process becomes more ergodic, as one can judge from Fig. \ref{fig-MSD_150}. In this limit the ensemble and time averaged MSD nearly coincide in the range of diffusion times we examined.  

\begin{figure}
\includegraphics[width=7cm]{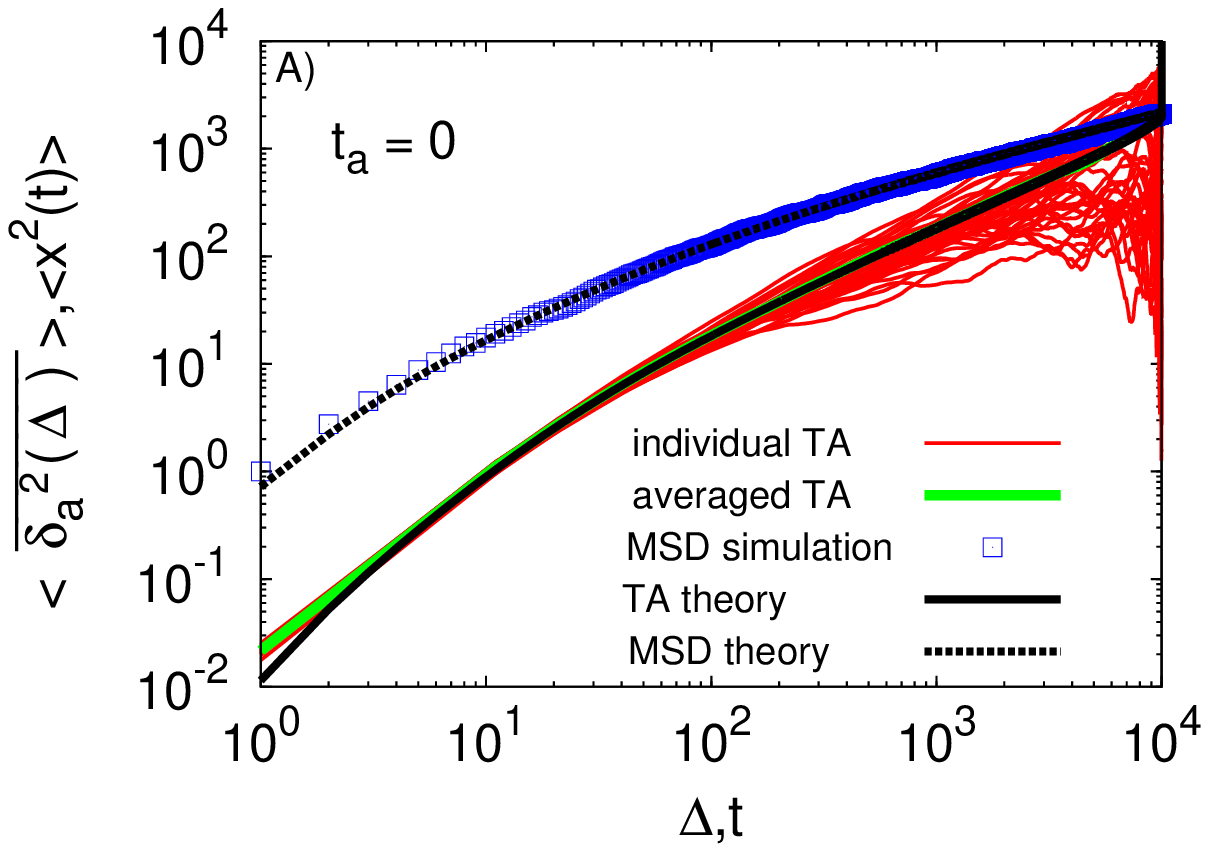}
\includegraphics[width=7cm]{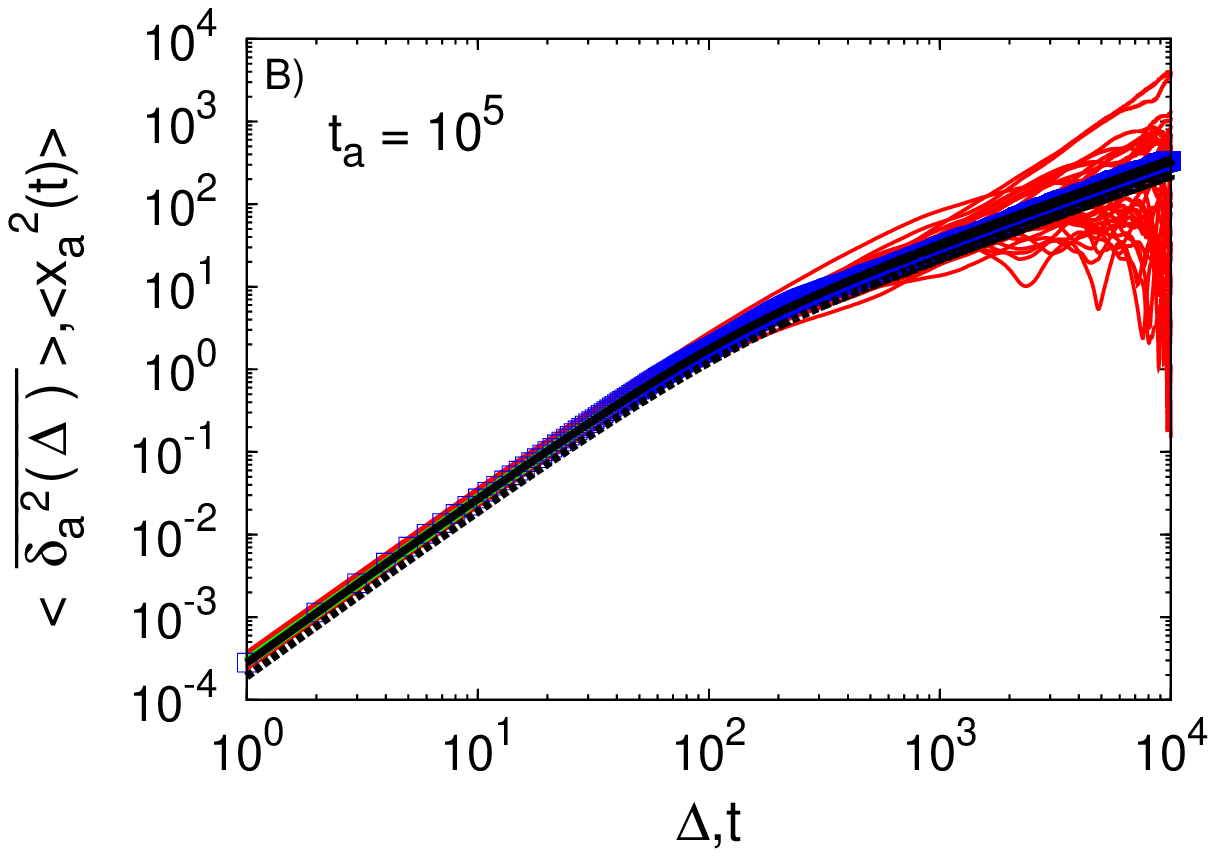}
\caption{MSD (blue points and dashed lines) and time averaged MSD (green lines and solid black line) for ageing UDSBM processes, for $\alpha=1/2$ and two different ageing times $t_a$ as indicated in panels A and B. The analytical expressions for the MSD and time averaged MSD are given by Eqs. (\ref{eq-msd-general-after-G-expansion}) and (\ref{eq-split}), respectively. The red curves represent individual time averaged MSD realizations. Other parameters are the same as in Fig. \ref{fig-msd}A.} \label{fig-tamsd-spread-all-graphs} \end{figure}

In the case of subdiffusion, see Fig. \ref{fig-tamsd-ballistic}, we expand the exponential function in Eq. (\ref{tamsd-SP1}) or Eq. (\ref{tamsd-SP4}) up to the second order in $\Delta$ and then integrate. The terms linear in the lag time from the main contribution (\ref{eq-tamsd-no-ageing-massive-delta0}) and from the additional term (\ref{tamsd-SP4}) vanish, while the second order in $\Delta$ produces for the time averaged MSD the initial ballistic regime \begin{equation} \left\langle\overline{ \delta^2_a(\Delta)} \right\rangle \sim D_0 \gamma_0 \left(\frac{t_a}{\tau_0}\right) ^{2\alpha-2} \Delta^2 \simeq \Delta^2. \label{eq-strong-ageing-limit-tamsd} \end{equation} This regime extends up to lag times \begin{equation} \Delta<\Delta_\text{min} \sim (t_a/\tau_0) ^{1-\alpha}/\gamma_0, \label{eq-delta-critical-tamsd-ballistic} \end{equation} that is much longer than the characteristic time $1/\gamma_0$ as $t_a/\tau_0$ is a large parameter. Therefore, comparing Eq. (\ref{ballisticMSD}) for the MSD and Eq. (\ref{eq-strong-ageing-limit-tamsd}) for the time averaged MSD one can conclude that the initial ballistic behavior of strongly ageing UDSBM processes is nearly ergodic. 
This important effect is illustrated in Fig. \ref{fig-tamsd-spread-all-graphs}B, which also shows how the ergodicity of the ageing UDSBM process is recovered in the limit of long ageing times.

\begin{figure}\includegraphics[width=7cm]{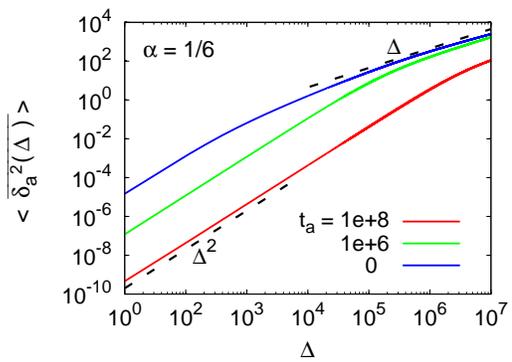}
\caption{Analytical results for the time averaged MSD of ageing UDSBM, plotted for $\alpha=1/6$ and varying ageing times. The asymptotes are according to Eqs. (\ref{eq-strong-ageing-limit-tamsd}) and (\ref{eq-tamsd-anna-eq17}). Parameters: $T=10^7$, $\tau_0=10^3$, and $\gamma_0=0.02$.} \label{fig-tamsd-asymp} \end{figure}

\begin{figure}\includegraphics[width=7cm]{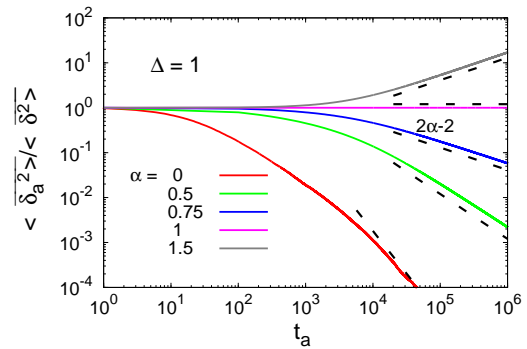}
\caption{Ratio of the ageing versus non-ageing time averaged MSDs of UDSBM, $\Lambda_\alpha(t_a)$,  as obtained from our computer simulations, plotted as a function of the ageing time $t_a$ for $\Delta=1$ and different values of the scaling exponents. Parameters: $T=10^4$, $N=10^3$. The analytical asymptotes in the limit of long ageing times  (\ref{eq-tamsd-strong}) are shown by the dashed lines.} \label{fig-tamsd-ratio} \end{figure}

The short lag time asymptote  (\ref{eq-strong-ageing-limit-tamsd}) and the full expression given by Eqs. (\ref{eq-tamsd-no-ageing-massive-delta0}), (\ref{tamsd-SP1}) are in a good agreement with the results of our numerical modeling of the Langevin equation, see Fig. \ref{fig-tamsd-ballistic}. In the regime of strong ageing when $t_a \gg T$, for strongly subdiffusive UDSBM the quadratic scaling of $\left\langle\overline{ \delta^2_a(\Delta)} \right\rangle$ with the lag time extends up to the entire observation period. This is another a priori surprising feature, rendering a purely overdamped description invalid. 

The effects of varying ageing time for the small value of  $\alpha=1/6$---relevant to the behavior of granular gases \cite{bodr15gg,bodr16}---can be seen in Fig. \ref{fig-tamsd-asymp} for rather long traces with $T=10^7$. When the ageing time is shorter than the observation time and the relation $T \gg \Delta \gg \tau_0$ is satisfied, the time averaged MSD has an extended intermediate linear scaling regime, in accord with the analytical prediction (\ref{eq-tamsd-anna-eq17}). This regime disappears in the limit of strong ageing by virtue of the fact that the initial ballistic regime for small $\alpha$ values extends to much longer times, in accord with Eq. (\ref{eq-delta-critical-tamsd-ballistic})---see also the curve for $t_a=10^8$ in Fig. \ref{fig-tamsd-ballistic}. This intermediate Fickean diffusion regime exists also for non-ageing UDSBM    \cite{bodr16}.

\subsubsection{Superdiffusion versus subdiffusion: weak ageing}
\label{sec-weak-ageing}

When the lag time $\Delta$ is the shortest time scale, we expand the contribution to the time averaged MSD $\left\langle\overline {\delta_{0,a} ^2(\Delta)}\right\rangle$ in Eq. (\ref{eq-tamsd-no-ageing-massive-delta0}) as well as the integrand of  $\left\langle \Xi_a(\Delta)\right\rangle$ in Eq. (\ref{tamsd-SP1}) for short lag times up to second order. Taking the integral in Eq. (\ref{tamsd-SP1}) and summing the two terms, we find that the contributions linear in $\Delta$ cancel and the leading order is quadratic in $\Delta$. This approximate expression for $\left\langle\overline{ \delta_{a}^2(\Delta)} \right\rangle$ at the conditions of weak ageing $t_a\ll T$ and in the physically relevant limit of   $\tau_0/T \ll 1$  for $0<\alpha<1/2$ has the form \begin{equation}\left\langle \overline{ \delta^2_a(\Delta)} \right\rangle \sim \frac{D_0 \gamma_0 \tau_0 }{(1-2 \alpha) T} \Delta^2, \label{eq-tamsd-general-small-delta} \end{equation} while at $\alpha>1/2$ the leading order is \begin{equation}\left\langle \overline{\delta^2_a (\Delta)} \right\rangle \sim \frac{D_0 \gamma_0 }{(2 \alpha-1) }\left(\frac{T} {\tau_0}\right)^{2\alpha-2} \Delta^2. \label{eq-tamsd-general-small-delta-2} \end{equation} At  $\alpha=1$ this approximate procedure yields \begin{equation} \left\langle\overline{ \delta^2_a(\Delta)}\right\rangle \sim D_0 \gamma_0 \Delta^2,\label{eq-alfa-1}\end{equation} as follows also from Eq. (\ref{eq-general-tamasd-3}). The critical value of $\alpha=1/2$ demarcates the boundary for different scalings of the $\left\langle \overline{\delta^2_a (\Delta)} \right\rangle$ prefactors with the trace length $T$ in the short time limits (\ref{eq-tamsd-general-small-delta}) and (\ref{eq-tamsd-general-small-delta-2}). At $\alpha=1/2$ the prefactor becomes a logarithmic rather than a power law function of the trace length, namely, \begin{equation} \left\langle\overline{ \delta^2_a(\Delta)} \right\rangle \sim \frac{ D_0 \tau_0\gamma_0 }{T} \log \left(1+\frac{T} {\tau_0}\right) \Delta^2 . \label{eq-tamsd-alfa05-log} \end{equation}

\subsubsection{Time averaged MSD enhancement/suppression function}

{Let us now consider the degree of enhancement or suppression of the time averaged MSD due to the presence of ageing  \cite{bark13} in the UDSBM process. It is quantified by the ratio of the ageing versus non-ageing time averaged MSD magnitudes \begin{align}\Lambda_\alpha \left({t_a,\Delta}\right) =
 \frac{\left\langle\overline{\delta^2_a(\Delta)}\right\rangle}
{\left\langle\overline{\delta^2(\Delta)}\right\rangle}\end{align} Using the relations (\ref{eq-strong-ageing-limit-tamsd}) and (\ref{eq-tamsd-general-small-delta}), (\ref{eq-tamsd-general-small-delta-2}), (\ref{eq-tamsd-alfa05-log}) for short lag times (corresponding to the ballistic regime), long particle trajectories  $\{\Delta,\tau_0 \} \ll T$ and strong ageing $t_a\gg T$ one gets the following asymptotic form \begin{align} \Lambda_\alpha \left({t_a}\right)  \sim \left\{ \begin{aligned}
        &(1-2 \alpha) \frac{T}{\tau_0}   \left(\frac{t_a}{\tau_0}\right)^{2\alpha-2}, ~&\alpha < 1/2\\         & \frac{{T}/{\tau_0}}{\log(1+T/\tau_0)}\left(\frac{t_a} {\tau_0}\right) ^{2\alpha-2},~&\alpha=1/2\\         &(2\alpha-1) \left(\frac{t_a}{T} \right)^ {2\alpha-2},~&\alpha > 1/2      \end{aligned}.\right.  \label{eq-tamsd-strong} \end{align} In this limit, the quadratic dependence of $\Lambda_\alpha$ on the lag time $\Delta$ in time averaged MSDs cancels out in Eq.  (\ref{eq-tamsd-strong}) and the universal  power law scaling in the leading order is \begin{equation} \Lambda_\alpha(t_a)\simeq t_a^{2\alpha-2}. \label{eq-lambda-alfa-scal} \end{equation} Note here that for the standard overdamped SBM process the suppression (enhancement) function $\Lambda_\alpha $ for subdiffusive (superdiffusive) realizations of the scaling exponent $\alpha$ is \cite{jeon15, cher15b} \begin{align} \label{eq-tamsd-ratio-overdamped} \Lambda_\alpha (t_a) \approx \left(1+\frac{t_a}{T}\right)^{\alpha} - \left(\frac{t_a}{T}\right)^ {\alpha}. \end{align} This power law function is similar to that observed for ageing continuous time random walks \cite{bark13} and ageing heterogeneous diffusion processes \cite{cher14b}.} 

{For ageing superdiffusive UDSBM processes, as predicted by Eq. (\ref{eq-tamsd-strong}) at  $\alpha>1$, the magnitude of the time averaged MSD gets enhanced with the ageing time, while for subdiffusive UDSBMs realized at $\alpha<1$ the time averaged MSD gets suppressed with increasing $t_a$. The analytical estimate (\ref{eq-tamsd-strong}) is in good agreement with our computer simulations of the ageing UDSBM process, as presented in Fig. \ref{fig-tamsd-ratio} for systematically varied exponent $\alpha$. We mention that the smaller the scaling exponent $\alpha$, the more pronounced is the decrease of the time averaged MSD with the ageing time $t_a$, while for $\alpha>1$ the enhancement of the time averaged MSD is observed, in accord with Eq. (\ref{eq-lambda-alfa-scal}). In Fig. \ref{fig-tamsd-spread-all-graphs} for a special value of $\alpha=1/2$ we also observe that the time averaged MSD gets reduced for longer ageing times. Finally note a different dependence of $\Lambda_\alpha(t_a)$ in Eq. (\ref{eq-tamsd-strong}) on the length of the particle trajectory $T$ for subdiffusive versus superdiffusive UDSBM processes, as well as for the critical value of the scaling exponent $\alpha=1/2$.

\begin{figure}
\includegraphics[width=7cm]{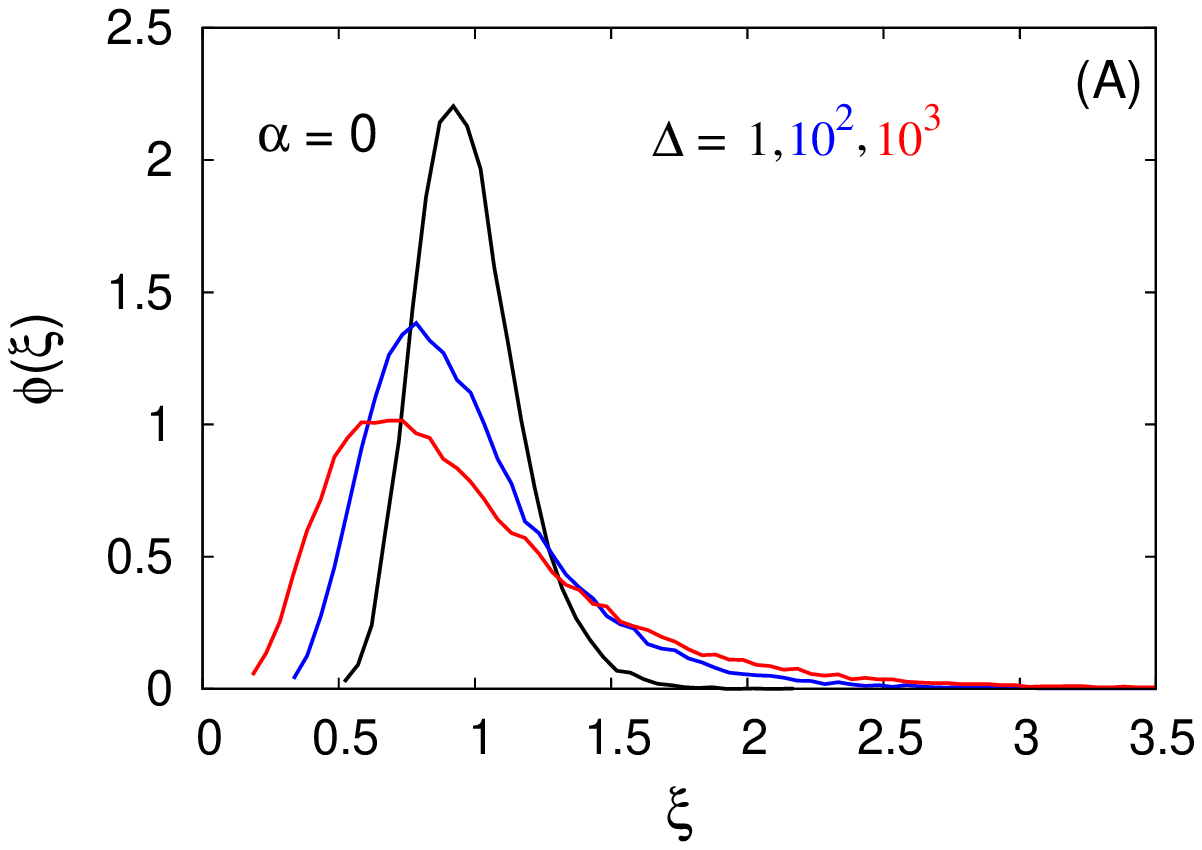}  
\includegraphics[width=7cm]{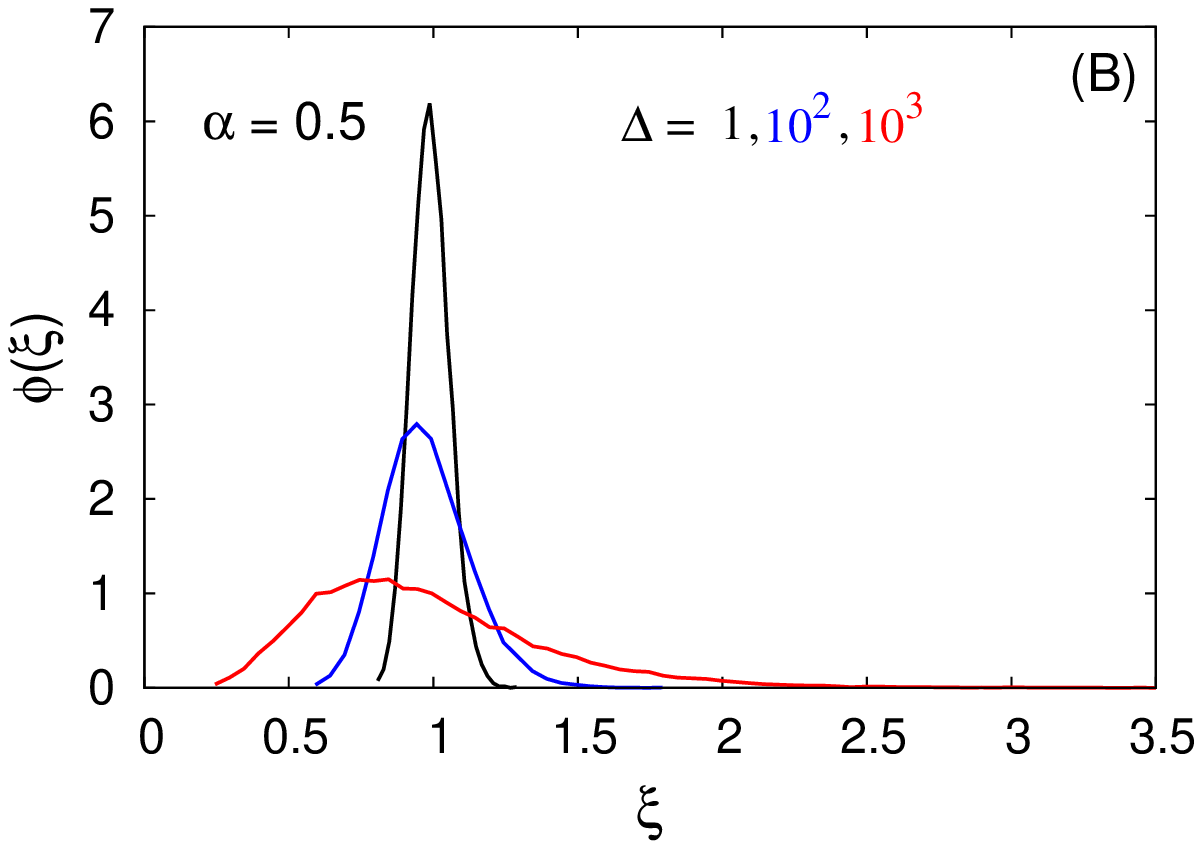}
\includegraphics[width=7cm]{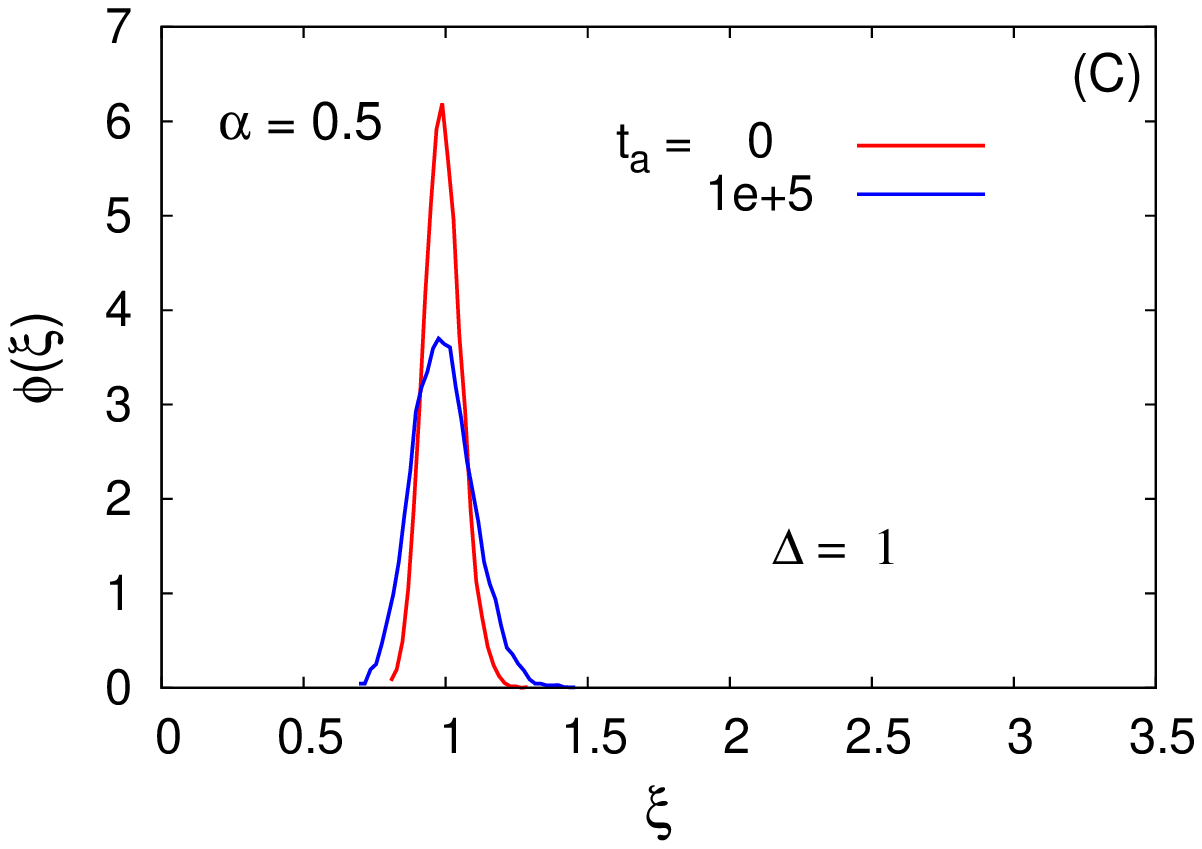} 
\includegraphics[width=7cm]{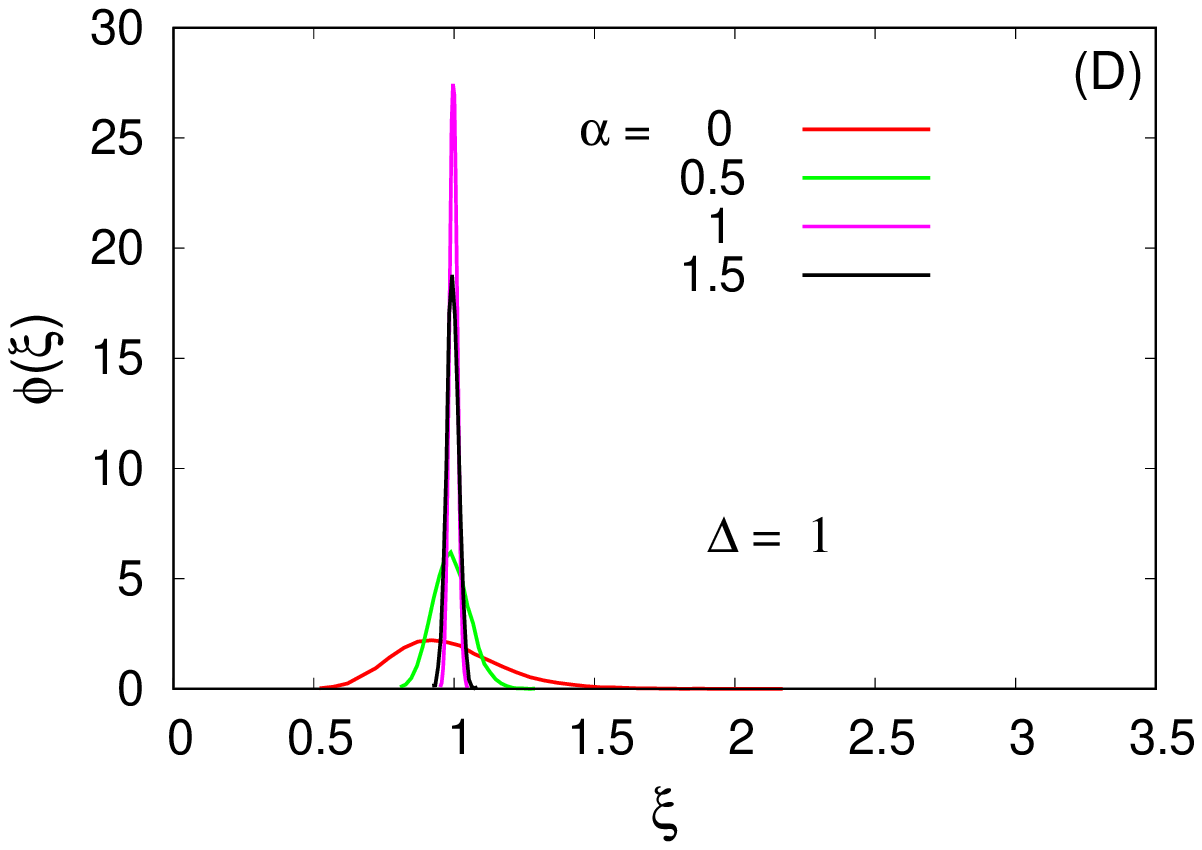} 
\caption{(A,B): Distribution $\phi(\xi)$ of the relative amplitude of the time averaged MSD for non-ageing UDSBM at $\alpha=1/2$ and ultraslow UDSBM, computed for different lag times, as indicated in the panels. For longer lag times the distribution gets progressively  wider and becomes asymmetric. (C): Comparison between $\phi(\xi)$ for ageing and non-ageing UDSBM processes for $\alpha=0.5$ and lag time $\Delta=1$. In the strong ageing regime the distribution becomes slightly wider. (D): Distributions $\phi(\xi)$ for different $\alpha$ exponents as indicated in the plot and for $\Delta=1$. Parameters for all the plots  are $T=10^{4}$ and  $N=10^{4}$.} \label{fig-PHI} \end{figure}

\begin{figure}\includegraphics[width=7cm]{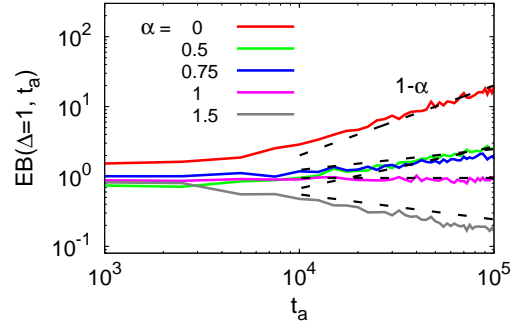}
\caption{EB parameter obtained from computer simulations versus the ageing time $t_a$ for the ageing UDSBM process, computed for traces with $T=10^{4}$ steps and averaged over $N=10^3$ trajectories, for different $\alpha$ values (as denoted in the plot) and $\Delta=1$. The black dashed lines show the phenomenological strong ageing scaling relation  (\ref{eq-eb-versus-t}).} \label{fig-eb}\end{figure} 

\subsection{Scatter of time averaged MSDs and ergodicity breaking parameter}

For a finite trajectory length all stochastic processes exhibit trajectory-to-trajectory fluctuations. These lead to fluctuating apparent mobilities of the particles. Figure \ref{fig-tamsd-spread-all-graphs} illustrates that the amplitude scatter of individual time averaged MSD trajectories of both ageing and non-ageing UDSBM processes. It appears to be quite narrow and thus the process is fairly reproducible. In particular, the spread of time averaged MSD trajectories does not change much with the ageing time. At short lag times $\Delta$ the spread is rather small, similar to that for the Brownian motion with the same trace length $T$ \cite{metz14}. 

Fig. \ref{fig-PHI} shows the amplitude scatter of individual time averaged MSD traces $\phi(\xi_{}),$ for both the ageing and non-ageing UDSBM processes, for a set of  values of the diffusion exponent $\alpha$, lag time $\Delta$, and ageing time $t_a$. We first start with UDSBM in the absence of ageing, complement the results published in Ref. \cite{bodr16}. As follows from Figs. \ref{fig-PHI}A,B evaluated at $t_a=0$, larger values of $\Delta$ lead to more asymmetric, non-centered $\phi(\xi)$ distributions. Comparing $\phi(\xi)$ for non-ageing ultraslow ($\alpha=0$) and subdiffusive ($\alpha=1/2$) UDSBM ---see panels A and B of Fig. \ref{fig-PHI}---we clearly see a broader spread of time averaged MSD realizations for $\alpha=0$ situation. Systematically larger values of the EB parameter found in simulations at $\alpha=0,$ as demonstrated in Fig. \ref{fig-eb}, are in line with this larger widths of $\phi(\xi)$ distributions for the ultraslow UDSBM process.

For fixed $\Delta$ and  $\alpha$ values, the presence of ageing in the system makes the distributions of the time averaged MSDs slightly wider, see Fig. \ref{fig-PHI}C. According to Fig. \ref{fig-PHI}D, by increasing the scaling exponent $\alpha$ up to unity the distributions $\phi(\xi)$ become narrower. For more superdiffusive $\alpha$ values however, the distribution $\phi(\xi)$ becomes slightly broader again. This effect is similar to the dependence of the ergodicity breaking parameter for the standard SBM process as a function of the scaling exponent (see the description and Fig. 3A in Ref. \cite{safd15}). 

A fairly reproducible behavior of time averaged MSDs---that is narrow $\phi(\xi)$ distributions observed here---is similar to that of otherwise ergodic fractional Brownian motion and fractional Langevin equation motion \cite{bark09, jeon10a, jeon10b}. Note that for these processes the effects of transient ageing and weak ergodicity breaking were also studied \cite{kurs13, jeon12age, jeon13age}. This reproducibility of time averaged MSDs for ageing UDSBM is in strong contrast, for instance, to continuous time random walks in which time averages of physical observables remain random quantities, even in the limit $T\to\infty$ \cite{metz14,soko08}.  

A commonly used measure of these amplitude fluctuations of time averaged MSD trajectories for UDSBM and other anomalous diffusion processes is the ergodicity breaking parameter, $\text{EB}$, defined via Eq. (\ref{eq-eb-general}) \cite{metz14}. In Fig. \ref{fig-eb} we present the variation of EB for the ageing UDSBM with the ageing time $t_a$, as obtained from our computer simulations. Performing a fitting to data points, we find that for strong ageing the following scaling is valid, \begin{equation} \text{EB}( t_a)\simeq t_a^{1-\alpha}. \label{eq-eb-versus-ta} \end{equation} Therefore, with increasing ageing time the EB parameter decreases for superdiffusive and grows for subdiffusive UDSBM. This is consistent with a somewhat broader distributions of the time averaged MSD traces obtained for superdiffusive UDSBMs in Fig. \ref{fig-PHI}D, as compared to the distribution $\phi(\xi)$ at $\alpha=1$. We note here that the number of traces needed to reach a satisfactory statistics for the EB parameter---containing the fourth moment of the time averaged MSDs---is typically considerably larger than that required for the convergence of  $\left<\overline{ \delta^2_a(\Delta)}\right>$  \cite{safd15}. For UDSBM we did not compute the ergodicity breaking parameter analytically, that might be a subject of a future study. 

The dependence of the $\text{EB}$ parameter on the trajectory length $T$ is presented in Fig. \ref{fig-eb-T}, for both ageing and non-ageing UDSBM. In the absence of ageing, as shown in Fig. \ref{fig-eb-T}A, $\text{the EB}$ parameter in the limit of long trajectories varies as \begin{equation} \label{eq-eb-versus-t} \text{EB}_\text{non-ageing} (T)\simeq 1/T^{\alpha}. \end{equation} Note that this exponent in this limit of short lag times for UDSBM is different from that for the standard SBM, where the dependence is EB$(T)\simeq1/T^{2\alpha}$ for $0<\alpha<1/2$ and EB$(T)\simeq1/T$ for $\alpha>1/2$ \cite{safd15}. However, in the limit of strong ageing the decay of EB is inversely proportionally with the trace length $T$, \begin{equation} \text{EB}_\text{ageing}(T)\simeq 1/T, \label{eq-eb-versus-t_aged} \end{equation} see Fig. \ref{fig-eb-T}B for a short lag times, $\Delta=1$. Note that the same scaling relation of EB is observed also for longer lag times (results not shown). Note also that this inverse proportionality of the EB parameter with the trace length is typical for a number of other anomalous diffusion processes, see Ref. \cite{metz14}.

\begin{figure*}
\includegraphics[width=15cm]{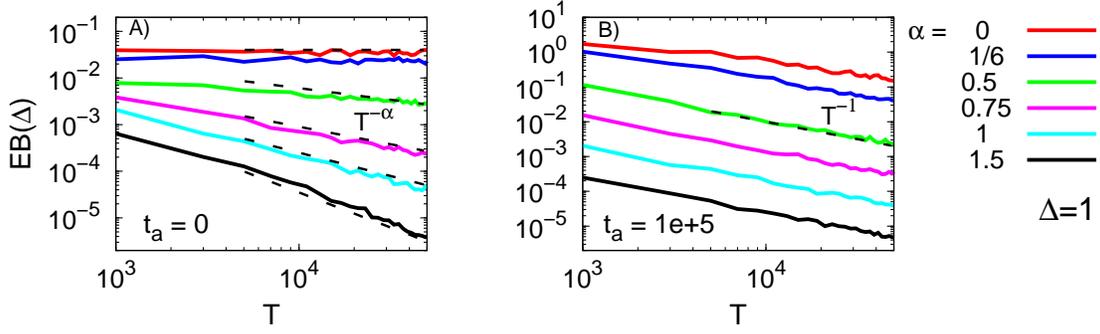}  
\caption{Ergodicity breaking parameter variation with the trace length $T$, computed for different $\alpha$ and two ageing regimes, after averaging over $N=10^{3}$ trajectories. The value of the lag time is $\Delta=1$.
The asymptotes shown in the plots are Eqs. (\ref{eq-eb-versus-t}) and (\ref{eq-eb-versus-t_aged}), in the limit of short and long ageing times, respectively.}
\label{fig-eb-T}\end{figure*}

\section{Main results: ageing ultraslow UDSBM}
\label{sec-uudsbm}

Let us now consider  inertial effects for the ageing UDSBM process at $\alpha=0$. The motion of massive particles in this case is governed by the underdamped Langevin equation \cite{bodr16}, \begin{align}  m\frac{d^{2}x(t)}{dt^{2}}  + \frac{\gamma_0}{\left(1+t/\tau_{0} \right)}\frac{dx(t)}{dt}=\sqrt{\frac{2D_0}{1+t/\tau_0}} \frac{\gamma_0 \eta(t)}{\left(1+t/\tau_{0} \right)}. \label{uusbm}\end{align} 

\subsection{MSD}

We straightforwardly obtain the velocity auto-correlation and find in the limit  $\tau_0\gamma_0\gg1$ that  \begin{equation} \left\langle v(t_1)v(t_2) \right\rangle \approx \frac{\mathcal{T}(0)} {m} \left(\frac{1+t_1/\tau_0}{1+t_2/\tau_0} \right)^{\tau_0\gamma_0} \frac{1} {(1+t_1/\tau_0)^2}. \end{equation} Note that this expression can be directly obtained from Eq. (\ref{v_corr}) by putting $\alpha\to0$. After some simplifications, the MSD of ageing ultraslow UDSBM process acquires both a logarithmic and power law function of the diffusion time $t$, \begin{align} \nonumber \left\langle x_a^2(t) \right\rangle \sim & 2D_0\tau_0 \log\left(1+\frac{t}{t_a+\tau_0}\right)
 \\  & +\frac{2D_0}{\gamma_0}\left[\left(1+\frac{t}{t_a+\tau_0}\right)^{-\tau_0\gamma_0} -1\right]. \label{R2age} \end{align} We restrict the analysis of this equation to the most interesting situation of strong ageing, \begin{equation} t_a\gg \{t,\tau_0\}. \end{equation} In this limit, Eq. (\ref{R2age}) can be approximated by  \begin{equation} \left\langle x_a^2(t)\right\rangle \sim 2D_0\tau_0 \frac{t}{t_a}- \frac{2D_0} {\gamma_0}\left(1-\exp\left[-\frac{t}{t_a}\tau_0\gamma_0\right]\right).
\label{R2age2} \end{equation}In the argument of the exponent we thus observe a product of a large parameter $\tau_0\gamma_0$ and a small parameter $t/t_a$. Therefore, for the range of diffusion times up to \begin{equation}t<t_\text{min}\sim t_a/(\tau_0\gamma_0)\end{equation} the leading order expansion of Eq. (\ref{R2age2}) the MSD of ageing ultraslow UDSBM---similarly to the MSD for the non-ageing UDSBM process
\cite{bodr16}---shows the ballistic regime,  \begin{equation}
\left\langle x_a^2(t)\right\rangle \sim \frac{D_0\tau_0^{2} \gamma_0 }{t_a^2} t^2 \simeq t^2. \label{eq-ultraslow-msd-ballistic} \end{equation} Note that this regime extends to the times much longer than the relaxation time $1/\gamma_0$ for standard diffusion \cite{uhle30} of massive Brownian particles, as can be seen from Fig. \ref{fig-UUDSBM}A. 

For the subsequent diffusion regime satisfying the condition \begin{equation}t>t_\text{min}\sim  t_{a}/(\tau_0\gamma_0),\end{equation} the first term in Eq. (\ref{R2age2}) dominates, yielding the linear growth of the MSD with time \begin{equation} \left\langle x_a^2(t)\right\rangle \sim 2D_0\tau_0  
\frac{t}{t_a} \simeq t. \label{eq-uudsbm-linear-msd-with-time} \end{equation} Similarly to  non-ageing UDSBM  \cite{bodr16}, for long observation times $t \gg \{t_a,\tau_0\}$  the MSD of ageing ultraslow UDSBM demonstrates (as expected) a logarithmic dependence on the diffusion time, \begin{equation} \left\langle x_a^2(t)\right\rangle \sim 2D_0 \tau_0\log\left(\frac{t} {t_a}\right). \label{eq-ulstraslow-msd-log}\end{equation}

\begin{figure}
\includegraphics[width=7cm]{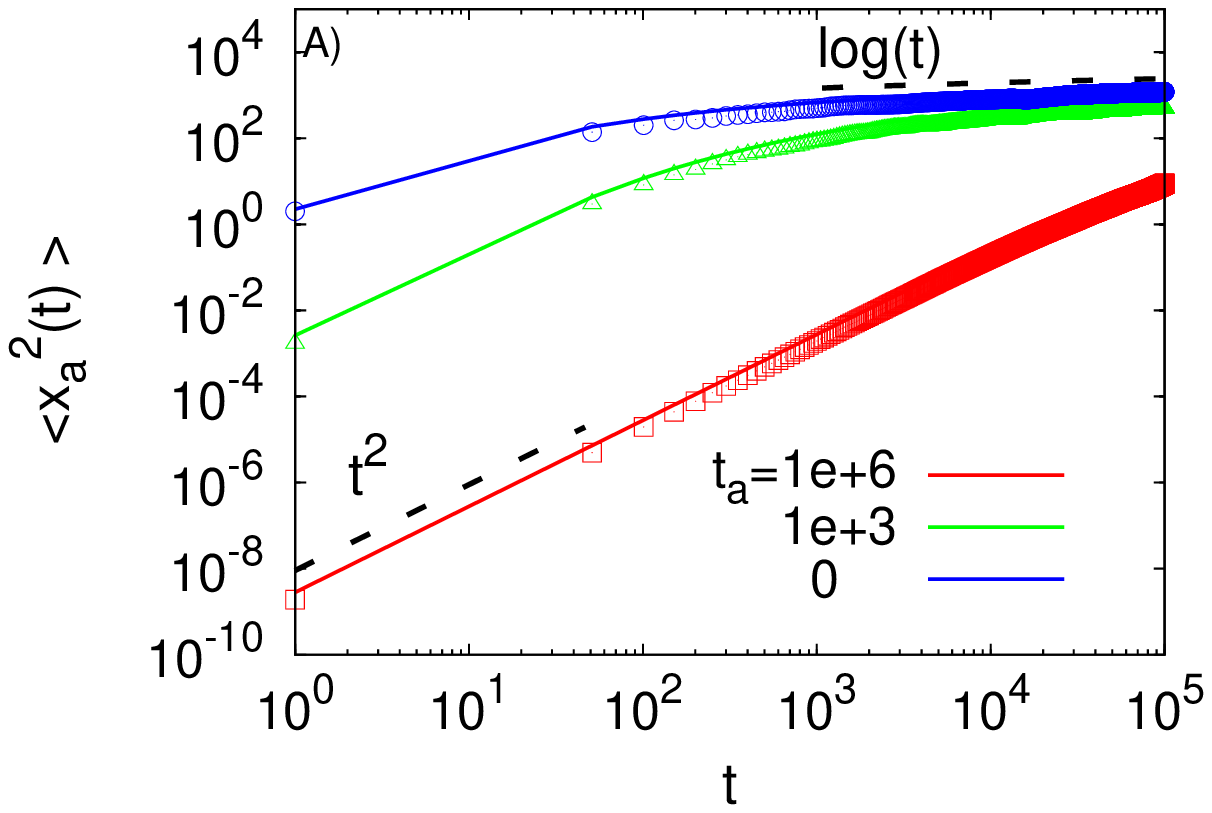} 
\includegraphics[width=7cm]{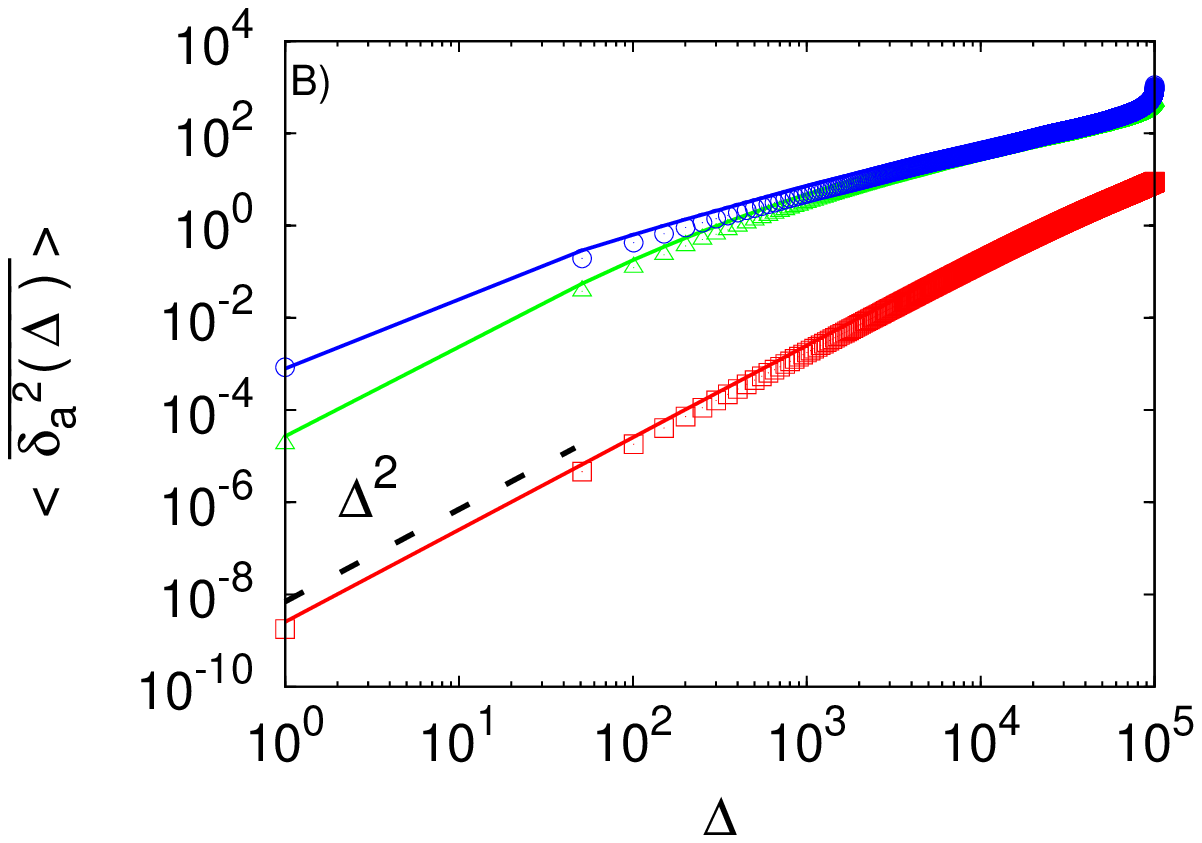}
\caption{MSD (panel A) and time averaged MSD (panel B) as obtained from computer simulations (data points) and theoretically (solid lines; {Eq. (\ref{R2age}) for the MSD and Eqs. (\ref{eq-tamsd-ages-ultra-zero-contribution}), (\ref{eq-theta-ultra}) for the time averaged MSD}) for the ageing ultraslow UDSBM process at $\alpha=0$.  The short time ballistic asymptote and the long time logarithmic behavior of the MSD are shown as dashed lines, plotted according to Eqs. (\ref{eq-ultraslow-msd-ballistic}) and (\ref{eq-ulstraslow-msd-log}), correspondingly. The ballistic asymptote for the time averaged MSD is the dashed lines given by Eq. (\ref{eq-ultraslow-tamsd-ballistic}). The values of the ageing time $t_a$ are as indicated in the plot. Other parameters are $D_0=1$, $\gamma_0=1$, $\tau_0=30,$ and $m=1$.}\label{fig-UUDSBM} \end{figure}

\subsection{Time averaged MSD}

The ageing time averaged MSD of the ultraslow UDSBM process acquires a form similar to Eq. (\ref{eq-split}), namely a combination of two terms. Here, the main contribution to the time averaged MSD coincides with that of the non-ageing ultraslow UDSBM process in the limit (\ref{eq-tau0-gamma0-big}), namely (see Eq. (61) in Ref. \cite{bodr16}) \begin{align} \nonumber 
\left< \overline{\delta_{0,a}^2 (\Delta)}\right>  \approx 
& \frac{2D_0\tau_0}{T-\Delta} \int_{t_a}^{T+t_a-\Delta}dt'\log \left(1+\frac{\Delta}{\tau_0+t'}\right)\\ \nonumber 
& = \frac{2D_0\tau_0}{T-\Delta}\left[ \left(T+t_a+\tau_0 \right) \log\left(1+\frac{T+t_a}{\tau_0}\right)  \right. \\ \nonumber 
& \left.- \left(t_a+\tau_0+\Delta\right)\log\left(1+\frac{t_a+\Delta}{\tau_0} \right)\right.   \\ \nonumber & -\left. \left(T+t_a+\tau_0-\Delta\right)\log\left(1+\frac{T+t_a- \Delta}{\tau_0}\right) \right. \nonumber  \\ 
& +\left. \left(t_a+\tau_0\right)\log\left(1+\frac{t_a} {\tau_0}\right)\right].  \label{eq-tamsd-ages-ultra-zero-contribution} \end{align} The second contribution to the time averaged MSD describes the inertial term in the original Langevin equation (\ref{uusbm}), namely, \begin{align} \nonumber \left< \Xi_a(\Delta)\right>\approx&\frac{2D_0} {\gamma_0(T-\Delta)} \int_{t_a}^{T+t_a- \Delta}dt^{\prime} \\ &\times  \left[\left(1+\frac{\Delta} {t^{\prime}+\tau_0}\right)^{-\tau_0\gamma_0}-1\right]. \label{eq-theta-ultra}\end{align} Using for Eq. (\ref{eq-theta-ultra}) the exponential representation analogous to that used for Eq. (\ref{tamsd-SP4}) in the limit of long ageing times, $t_a\gg \{T, \Delta\},$ we get the initial ballistic regime of the time average MSD, \begin{eqnarray} \left\langle\overline{\delta_a^2(\Delta )}\right\rangle \sim D_0\gamma_0 \left(\frac{\tau_0}{t_a}\right)^2 \Delta^2 \simeq\Delta^2. \label{eq-ultraslow-tamsd-ballistic} \end{eqnarray} Note that this relation can be also directly obtained from Eq. (\ref{eq-strong-ageing-limit-tamsd}) by putting $\alpha\to0$. The ballistic regime (\ref{eq-ultraslow-tamsd-ballistic}) extends up to lag times satisfying the condition \begin{equation}\Delta<\Delta_\text{min} \sim  t_a/(\tau_0\gamma_0),\end{equation} that is much longer that the relaxation time, $1/\gamma_0$. We remind the reader that the latter defines the time scale of the ballistic regime for ordinary Brownian motion, Eq. (\ref{eq-d0-bm}). Note that the effective diffusion constant for the ageing ultraslow UDSBM process, \begin{equation}D_\text{eff}(t_a)=D_0\gamma_0 \left(\frac{\tau_0}{t_a} \right)^2,\label{eq-deff-ultraslow} \end{equation}  in this limit of strong ageing becomes much smaller than for normal Brownian motion, $D_0\gamma_0$. Figure \ref{fig-UUDSBM} demonstrates a good agreement between our computer simulations of the underdamped equation (\ref{uusbm}) for the ageing ultraslow UDSBM process and the theoretical results, for both the MSD and the time averaged MSD. Finally, the scatter of time averaged MSDs for the ultraslow UDSBM process is illustrated in Fig. \ref{fig-PHI}A.
 
\section{Discussion and conclusions}
\label{sec-disc}

In the current study, we rationalized the effects of ageing on the ensemble averaged MSD, the time averaged MSD, and the ergodic properties of the underdamped SBM process (UDSBM). We explicitly considered the effects of a \textit{finite particle mass }on the magnitude and duration of the short time ballistic regime, both for the MSD and time averaged MSD. The thorough investigation of the effects of ageing on the UDSBM process complements and completes our recent study \cite{bodr16}. Ageing is shown to reduce the magnitude of the time averaged MSD for the case of subdiffusion and to increase it in the superdiffusion case. We showed also that for longer ageing times in the ballistic regime the MSD converges to the time averaged MSD and thus ergodicity is restored for the ageing UDSBM process. 

The existence and unexpectedly long persistence of the short time ballistic regime for the UDSBM process was first predicted in Ref. \cite{bodr16}. In the presence of ageing, however, the duration of this regime depends on the anomalous exponent as well as the ageing time. For subdiffusive exponents both the MSD and the time averaged MSD show a ballistic behavior for times considerably longer than that for ordinary Brownian motion,  $1/\gamma_0$. At later times, a transition to normal diffusion is observed for the ageing and non-ageing UDSBM processes. In contrast, for superdiffusive exponents $\alpha>1$ the normal diffusive regime dominates in the intermediate and long time regime. Our analytical results are supported by the findings of extensive computer simulations of the stochastic Langevin equation for massive particles in a medium with time varying diffusion coefficient. 

We characterized the behavior of the system both for subdiffusive and superdiffusive realizations of the scaling exponent $\alpha$, as well as for the limiting value of $\alpha=0$. The latter gives rise to ultraslow UDSBM, with a characteristic combined logarithmic and power law time dependence of the averaged particle displacement. Particularly for subdiffusive UDSBM processes we demonstrated that in the limit of long ageing times the overdamping approximation fails entirely. Instead, the initial ballistic regime expected for massive particles extends for large $t_a$ values up to the entire trace length, i.e. times much longer than typical relaxation time $1/\gamma_0$.

We also analyzed the non-ergodicity of the ageing UDSBM process based on the ergodicity breaking parameter, EB. Based on computer simulations, we demonstrated that the presence of ageing changes the decay  of EB. Specifically, for long observation times the EB of the UDSBM process in the limit of strong ageing tends to zero as $T^{-1}$, different from its $T^{-\alpha}$ asymptotic scaling for the non-ageing UDSBM process. Our findings support the idea that in non-stationary diffusive systems the presence of ageing can drastically alter the particle dynamics, at initial, intermediate, and long time limits. Additionally, the current study extends the range of scaling behaviors predicted for the non-ageing UDSBM process in Ref. \cite{bodr16}.

The applications of our results to real physical systems include the behavior and dynamics of particles in granular gases, with the power law decrease of the medium temperature \cite{Brilliantov, bodr15gg, bodr16}. It will be interesting to compare our results to more detailed simulations of these systems. The development of underdamped particle dynamics and approaches for other anomalous diffusion processes is also of great interest. For instance, the limits of applicability of the commonly used overdamping approximation for continuous time random walks---known to be connected to SBM in a mean field sense \cite{soko14}---would be intriguing to unravel in the future.






\end{document}